\documentclass[prd,aps,nofootinbib]{revtex4}
\usepackage{amssymb,graphicx}

\newtheorem{theorem}{Theorem}

\newtheorem{lemma}{Lemma}
\newtheorem{remark}{Remark}

\newcommand{\proof}{\noindent {\bf Proof. }}

\newcommand{\qed}{\hfill $\fbox{\hspace{0.3mm}}$ \vspace{.3cm}} 

\newcommand{\diag}{\mbox{diag}}

\newcommand{\Real}{\mathbb{R}}

\newcommand{\gz}{\mbox{\em \r{g}\hspace{0.3mm}}}  
\newcommand{\Rz}{\mbox{\em \r{R}}}

\newcommand{\nablaz}{\nabla\hspace{-0.27cm}{}^{\mbox{\r{~}}}{}\hspace{-0.22cm}}

\newcommand{\Gammaz}{\Gamma\hspace{-0.25cm}{}^{\mbox{\r{~}}}{}\hspace{-0.12cm}}

\begin{document}

\title{Boundary conditions for coupled quasilinear wave equations with
application to isolated systems}

\author{H.-O. Kreiss${}^{1,2}$, O. Reula${}^{3}$, O. Sarbach${}^{4}$
and J. Winicour${}^{2,5}$}

\affiliation{${}^{1}$NADA, Royal Institute of Technology, 10044
Stockholm, Sweden}

\affiliation{${}^{2}$Albert Einstein Institute, Max Planck
Gesellschaft, Am M\"uhlenberg 1, D-14476 Golm, Germany}

\affiliation{${}^{3}$FaMAF Universidad Nacional de Cordoba, Cordoba,
Argentina 5000}

\affiliation{${}^{4}$Instituto de F{\'{\i}}sica y Matem{\'{a}}ticas,
Universidad Michoacana de San Nicol{\'{a}}s de Hidalgo,
Edificio C-3, C.~P. 58040 Morelia, Michoac{\'{a}}n, M{\'{e}}xico}

\affiliation{${}^{5}$Department of Physics and Astronomy,
University of Pittsburgh, Pittsburgh, Pennsylvania 15260, USA}

\date{\today}

\begin{abstract}
We consider the initial-boundary value problem for systems of
quasilinear wave equations on domains of the form $[0,T] \times
\Sigma$, where $\Sigma$ is a compact manifold with smooth boundaries
$\partial\Sigma$. By using an appropriate reduction to a first order
symmetric hyperbolic system with maximal dissipative boundary
conditions, well posedness of such problems is established for a large
class of boundary conditions on $\partial\Sigma$. We show that our
class of boundary conditions is sufficiently general to allow for a
well posed formulation for different wave problems in the presence of
constraints and artificial, nonreflecting boundaries, including
Maxwell's equations in the Lorentz gauge and Einstein's gravitational
equations in harmonic coordinates. Our results should also be useful
for obtaining stable finite-difference discretizations for such
problems.
\end{abstract}

\maketitle

\section{Introduction and main results}
\label{Sect:Intro}

Motivated in part by the numerical computation of spacetimes on a
finite domain with artificial boundaries, the initial-boundary value
problem (IBVP) in general relativity has started to receive a lot of
attention during the last few years (see \cite{oS07} for a review). A
well posed IBVP for Einstein's vacuum field equations was formulated
for the first time by Friedrich and Nagy \cite{hFgN99} based on tetrad
fields and the theory of quasilinear, symmetric hyperbolic systems
with maximal dissipative boundary conditions
\cite{kF58,pLrP60,pS96b}. More recently, Kreiss and Winicour
\cite{hKjW06} formulated a well posed IBVP for the harmonic gauge
formulation of the Einstein vacuum equations which casts the field
equations into a set of ten coupled quasilinear wave equations subject
to four constraints. There are two key ideas behind the result of
\cite{hKjW06}. The first one is the realization that the wave
equations, when viewed as first order pseudodifferential equations,
has a non-characteristic boundary matrix. This allows application of
the boundary value theory for such systems developed by Kreiss in the
1970's \cite{hK70}. Second, the formulation of boundary conditions for
the frozen coefficient form of the harmonic Einstein equations which
ensure constraint propagation and satisfy the estimates required by
the Kreiss theory. The well posedness of the system and the
generalization to the quasilinear case can then be established using
the theory of pseudodifferential operators (see, for instance,
\cite{Taylor-Book2}).

In a subsequent paper \cite{hKoRoSjW07}, similar results were obtained
via more mundane energy estimates which follow by integration by
parts, without resort to the pseudodifferential calculus. For this, a
non-standard energy norm is constructed which is based upon the choice
of a particular time-like direction adapted to the boundary conditions
being imposed. With respect to this energy the Kreiss-Winicour
boundary conditions are maximally dissipative and so standard
well posedness theorems apply even in the quasilinear case
\cite{jRfM874,pS96b}. Besides being a simpler proof, or at least a
proof that can be followed completely by a reader not familiar with
the pseudodifferential techniques, it implies similar results for the
stability of finite difference approximations to Einstein's equations
in the harmonic gauge. This follows from considering the semidiscrete
system of ordinary differential equations in time obtained by
substituting finite differences for spatial derivatives. If the
semidiscrete system is stable, then for appropriate time
discretizations the fully discrete system is guaranteed to be
stable~\cite{KreisWu}. The stability of the semidiscrete system can be
established by the use of finite difference operators satisfying
summation by parts~\cite{KreisSch}, the counterpart of integration by
parts, by mimicking the steps leading to the continuum energy
estimate. A summation by parts algorithm for the harmonic Einstein
IBVP was developed for homogeneous boundary conditions \cite{mBbSjW06}
and verified to be stable in numerical tests~\cite{mBhKjW07}. The
results of \cite{hKoRoSjW07} provide a means to prove stability in the
inhomogeneous case.

In this paper we present a more general and geometric version of the
foregoing results which applies to coupled systems of quasilinear wave
equations with a certain class of boundary conditions. The
well posedness of the resulting IBVP is established by reducing the
wave system to first order symmetric hyperbolic equations subject to
maximally dissipative boundary conditions. As we show, our class of
boundary conditions is sufficiently flexible for obtaining well posed
IBVP formulations for different models of isolated systems in physics,
including the wave equation, Maxwell's equations and the Einstein
field equations.

In what follows we present the main results.

\subsection{Main theorem}

Let $T > 0$, and denote by $\Sigma$ a $d$-dimensional compact manifold
with smooth boundaries $\partial\Sigma$. The type of system our
results apply to is a set of quasilinear wave equations on $M = [0,T]
\times \Sigma$ coupled both by lower order terms and in the principal
part, by a change in the characteristic directions via a metric which
can depend on the local value of the fields involved. More precisely,
let $\pi: E \to M$ be a vector bundle over $M$ with fibre $\Real^N$,
let $\nabla_a$ be a fixed, given connection on $E$ and let $g_{ab} =
g_{ab}(\Phi)$ be a Lorentz metric on $M$ with inverse $g^{ab}(\Phi)$
which depends pointwise and smoothly on a set of fields $\Phi = \{
\Phi^A \}_{A=1,2,...N}$ parameterizing a local section of $E$. Our
signature convention for $g_{ab}$ is $(-,+,...,+)$. We shall also
assume that each time-slice $\Sigma_t = \{ t \} \times \Sigma$ is
space-like and that the boundary ${\cal T} = [0,T] \times
\partial\Sigma$ is time-like with respect to $g_{ab}(\Phi)$. In the
following, we will refer to local sections in $E$ as vector-valued
functions over $M$. We will also assume the existence of a
positive-definite fibre metric $h_{AB}$ on $E$. We consider a system
of quasilinear wave equations of the form
\begin{equation}
g^{ab}(\Phi) \nabla_a\nabla_b\Phi^A = S^A(\Phi,\nabla\Phi), 
\label{Eq:WaveSystemEq}
\end{equation}
where $S^A(\Phi,\nabla\Phi)$ is a vector-valued function which depends
pointwise and smoothly on its arguments. The wave system
(\ref{Eq:WaveSystemEq}) is subject to the initial conditions
\begin{equation}
\left. \Phi^A \right|_{\Sigma_0} = \Phi^A_0\; , \qquad
\left. n^b\nabla_b\Phi^A \right|_{\Sigma_0} = \Pi^A_0\; ,
\label{Eq:WaveSystemID}
\end{equation}
where $\Phi^A_0$ and $\Pi^A_0$ are given vector-valued functions on
$\Sigma_0$, and where $n^b = n^b(\Phi)$ denotes the future-directed
unit normal to $\Sigma_0$ with respect to $g_{ab}$. In order to
describe the boundary conditions, let $T^a = T^a(p,\Phi)$ be a
future-directed vector field on ${\cal T}$ which is normalized with
respect to $g_{ab}$ and let $N^a = N^a(p,\Phi)$ be the unit outward
normal to ${\cal T}$ with respect to the metric $g_{ab}$. We consider
boundary conditions on ${\cal T}$ of the following form\footnote{We
adopt the Einstein summation convention for the lower case Latin
abstract spacetime indices $a$, $b$, $c$, ... as well as for the
Capital indices $A$, $B$, $C$, ... on the fibre of $E$.}:
\begin{equation}
\left. \left[ T^b + \alpha N^b \right]\nabla_b\Phi^A \right|_{\cal T}
 = c^{a\, A}{}_B\left. \nabla_a\Phi^B \right|_{\cal T} 
 + d^A{}_B\left. \Phi^B \right|_{\cal T} + G^A,
\label{Eq:WaveSystemBC}
\end{equation}
where $\alpha = \alpha(p,\Phi) > 0$ is a strictly positive, smooth
function, $G^A = G^A(p)$ is a given, vector-valued function on ${\cal
T}$ and the matrix coefficients $c^{a\, A}{}_B = c^{a\,
A}{}_B(p,\Phi)$ and $d^A{}_B = d^A{}_B(p,\Phi)$ are smooth functions
of their arguments. Furthermore, we assume that $c^{a\, A}{}_B$ can be
made arbitrarily small in the following sense: Given a local
trivialization $\varphi: U \times \Real^N \mapsto \pi^{-1}(U)$ of $E$
such that $\bar{U}\subset M$ is compact and contains a portion ${\cal
U}$ of the boundary ${\cal T}$, and given $\varepsilon > 0$, there
exists a smooth map $J: U \to GL(N,\Real), p \mapsto (J^A{}_B(p))$
such that the transformed matrix coefficients
\begin{displaymath}
\tilde{c}^{a\, A}{}_B := J^A{}_C c^{a\, C}{}_D \left( J^{-1}\right)^D{}_B
\end{displaymath}
satisfy the condition
\begin{equation}
h_{AB}\tilde{c}^{a\, A}{}_C(\Phi)
      \tilde{c}^{b\, B}{}_D(\Phi) V_a{}^C V_b{}^D 
 \leq \varepsilon h_{AB} e^{ab}(\Phi) V_a{}^A V_b{}^B\; ,
\label{Eq:hABCondition}
\end{equation}
for all vector-valued one-forms $V_a^A$ on ${\cal U}$, where here and
in the following, $e_{ab}$ refers to the Euclidean metric $e_{ab} =
g_{ab} + 2T_a T_b$ which is defined for points on ${\cal T}$.

The main result of this paper is:

\begin{theorem}
\label{Thm:MainTheorem}
The IBVP
(\ref{Eq:WaveSystemEq},\ref{Eq:WaveSystemID},\ref{Eq:WaveSystemBC}) is
well posed. Given $T > 0$ and sufficiently small and smooth initial
and boundary data $\Phi_0^A$, $\Pi_0^A$ and $G^A$ satisfying the usual
compatibility conditions at $\partial\Sigma_0$, there exists a unique
smooth solution on $M$ satisfying the evolution equation
(\ref{Eq:WaveSystemEq}), the initial condition (\ref{Eq:WaveSystemID})
and the boundary condition (\ref{Eq:WaveSystemBC}). Furthermore, the
solution depends continuously on the initial and boundary data.
\end{theorem}

A common situation in which the condition (\ref{Eq:hABCondition}) is
automatically satisfied is given in the following

\begin{lemma}
\label{Lem:LyapunovTrick}
Let ${\cal U} \subset {\cal T}$ be an open subset of ${\cal T}$ such
that $\bar{U}$ is compact. Assume there exists a smooth map $J: {\cal
U} \to GL(N,\Real), p \mapsto (J^A{}_B(p))$ over ${\cal U}$ such that
the transformed matrix coefficients $\tilde{c}^{a\, A}{}_B := J^A{}_C
c^{a\, C}{}_D \left( J^{-1}\right)^D{}_B$ are in upper triangular form
with zeroes on the diagonal, that is
\begin{displaymath}
\tilde{c}^{a\, A}{}_B = 0,\qquad B \leq A.
\end{displaymath}
Then, the condition (\ref{Eq:hABCondition}) is satisfied on ${\cal U}$.
\end{lemma}

\proof (cf. The proof of the Liapunov stability theorem) In order to
simplify the notation we use a matrix notation and write $\tilde{c}^a
= J c^a J^{-1}$. Let $\delta > 0$, and define $D_\delta
:=\diag(1,\delta,\delta^2,...,\delta^{N-1})$ and $J_\delta
:=D_\delta^{-1} J$. Then, $c^a_\delta := J_\delta c^a J_\delta^{-1}
=D_\delta^{-1} \tilde{c}^a D_\delta$ has the components
$(c^a_\delta)^{A}{}_B = \delta^{B-A}\tilde{c}^{a\, A}{}_B$, where
here, $\delta^{B-A}$ refers to the $(B-A)$'th power of $\delta$. Since
$\tilde{c}^{a\, A}{}_B = 0$ for $B \leq A$ we have $c^a_\delta = {\cal
O}(\delta)$, and $c^a_\delta$ satisfies the condition
(\ref{Eq:hABCondition}) provided $\delta > 0$ is chosen small enough.
\qed

The proof of theorem \ref{Thm:MainTheorem} is given in sections
\ref{Sect:SimpleWave} and \ref{Sect:WaveSystems}. In order to
illustrate the ideas on a simpler example, we start in
Sect.~\ref{Sect:SimpleWave} with the wave equation on a {\em fixed}
background metric $g_{ab}$, and analyze the general case in
Sect.~\ref{Sect:WaveSystems}.

Since many physical systems can be described by systems of wave
equations, theorem \ref{Thm:MainTheorem} should have many
applications. In the following, we mention two such applications for
the initial-boundary value formulation of isolated systems with
constraints. The physical motivation for the choice of nonreflecting
boundary conditions in these examples is described in detail in
section \ref{Sect:Applications}.

\subsection{Maxwell's equations in the Lorentz gauge}

The first application describes an electromagnetic field on the
manifold $M = [0,T] \times \Sigma$ with a fixed background metric
$g_{ab}$ and corresponding Levi-Civita connection $\nabla_a$. As
before, we assume that each time-slice $\Sigma_t = \{ t \} \times
\Sigma$ is space-like and that the boundary ${\cal T} = [0,T] \times
\partial\Sigma$ is time-like. In the Lorentz gauge $C := \nabla_b A^b
= 0$, where $A^b$ denotes the $4$-vector potential, Maxwell's
equations assume the form of a system of wave equations,
\begin{equation}
g^{ab}\nabla_a\nabla_b A^c = R^c{}_d A^c - J^c\; ,
\label{Eq:MaxwellWave}
\end{equation}
where $R_{ab}$ denotes the Ricci tensor belonging to the metric
$g_{ab}$ and $J^c$ is the four-current. (\ref{Eq:MaxwellWave}) implies
that the constraint variable $C$ obeys the following equation
\begin{equation}
g^{ab}\nabla_a\nabla_b C = -\nabla^c J_c\; .
\label{Eq:PropConsMax}
\end{equation}
Therefore, the imposition of the boundary condition $\left. C
\right|_{\cal T} = 0$ and the satisfaction of the continuity equation
$\nabla^c J_c = 0$ imply that any smooth enough solution of
(\ref{Eq:MaxwellWave}) with initial data satisfying
\begin{displaymath}
\left. C \right|_{\Sigma_0} = 0, \qquad
\left. n^a\nabla_a C \right|_{\Sigma_0} = 0,
\end{displaymath}
satisfies the constraint $C=0$ on $M$ since in this case the
constraint propagation system (\ref{Eq:PropConsMax}) is homogeneous.

Asymptotically nonreflecting boundary conditions at ${\cal T} =
[0,T]\times\Sigma$, in the sense of \ref{Sect:Applications}, can be
formulated by first introducing a null tetrad $\{
K^a,L^a,Q^a,\bar{Q}^a \}$ which is adapted to the boundary. Let $T^a$
be a future-directed time-like vector field tangent to ${\cal T}$
normalized such that $g_{ab} T^a T^b = -1$, let $N^a$ denote the unit
outward normal to ${\cal T}$ with respect to $g_{ab}$ and complete
$T^a$ and $N^a$ to an orthonormal basis $\{ T^a,N^a,V^a,W^a \}$ of
$T_p M$ at each point $p\in{\cal T}$. Then, we define the null vectors
\begin{displaymath}
K^a := T^a + N^a, \qquad
L^a := T^a - N^a, \qquad
Q^a := V^a + i\, W^a, \qquad
\bar{Q}^a := V^a - i\, W^a,
\end{displaymath}
where $i = \sqrt{-1}$. These vectors may be smoothly continued in a
small region inside the domain, for example by parallel transport
along the normal direction to the boundary. In this way, one obtains a
local null basis of $TM$. Finally, let $r$ denote the areal radius of
the cross sections $\partial\Sigma_t$. This function can also be
continued in a small region inside the domain by parallel transporting
$\partial\Sigma_t$ along the normal direction. The following boundary
conditions are derived in section \ref{SubSec:NRBCMaxwell}
\begin{eqnarray}
\left. \frac{1}{r^2} K^a K_b \nabla_a(r^2 A^b) \right|_{\cal T} 
 &=& q_K\; ,
\label{Eq:MaxwellBC1}\\
\left. \left( K^a Q_b - Q^a K_b \right)\nabla_a A^b \right|_{\cal T}
 &=& q_Q\; ,
\label{Eq:MaxwellBC2}\\
\left. \left( K^a L_b + L^a K_b \nabla_a A^b 
            - Q^a\bar{Q}_b - \bar{Q}_a Q^b \right)\nabla_a A^b 
 \right|_{\cal T} &=& 0,
\label{Eq:MaxwellBC3}
\end{eqnarray}
where $q_K$ and $q_Q$ are given real and complex scalars on ${\cal
T}$. The first condition is a gauge condition, the second condition
controls the electromagnetic radiation through ${\cal T}$ and the
third condition enforces the constraint $C = g^{ab}\nabla_a A_b = 0$
on ${\cal T}$.

The evolution equation (\ref{Eq:MaxwellWave}) has the form
(\ref{Eq:WaveSystemEq}) where $E$ is the tangent bundle over $M$, and
the boundary conditions
(\ref{Eq:MaxwellBC1},\ref{Eq:MaxwellBC2},\ref{Eq:MaxwellBC3}) have the
form (\ref{Eq:WaveSystemBC}) with
\begin{eqnarray}
&& \alpha = 1,
\nonumber\\
&& c^{a\, c}{}_d = \frac{1}{2}\left[ 2 Q^{(a} \bar{Q}^{c)} K_d 
 + L^a K^c K_d - K^c\left( Q^a\bar{Q}_d + \bar{Q}^a Q_d \right) \right],
\qquad
d^c{}_d = (K^b\nabla_b\log r) L^c K_d\; , 
\nonumber\\
&& G^c = \frac{1}{2}
         \left[ -L^c q_K + \bar{Q}^c q_Q + Q^c\bar{q}_Q \right].
\nonumber
\end{eqnarray}
Since
\begin{eqnarray}
&& c^{a\, c}{}_d K^d = 0,
\nonumber\\
&& c^{a\, c}{}_d Q^d = -Q^a K^c,
\nonumber\\
&& c^{a\, c}{}_d\bar{Q}^d = -\bar{Q}^a K^c,
\nonumber\\
&& c^{a\, c}{}_d L^d = -L^a K^c - \bar{Q}^a Q^c - Q^a\bar{Q}^c,
\nonumber
\end{eqnarray}
the matrix elements $c^{a\, c}{}_d$ are in upper triangular form with
zeroes in the diagonal when expressed in terms of the basis $\{
K^a,Q^a,\bar{Q}^a,L^d \}$. Therefore, the assumptions of Lemma
\ref{Lem:LyapunovTrick} are satisfied and we obtain a well posed IBVP.

\subsection{Einstein's equations in harmonic coordinates}
\label{SubSec:Harmbg}

As a second application of our theorem we consider Einstein's field
equations in (generalized) harmonic coordinates. For this, we follow
\cite{HawkingEllis-Book,mRoRoS07} and choose a {\em fixed} background
metric $\gz_{ab}$ on $M = [0,T] \times \Sigma$ with the property that
each time-slice $\Sigma_t = \{ t \} \times \Sigma$ is space-like and
the boundary ${\cal T} = [0,T] \times \partial\Sigma$ is time-like
with respect to $\gz_{ab}$. We impose the following gauge condition on
the dynamical metric $g_{ab}$,
\begin{equation}
{\cal C}^c := g^{ab}\left( \Gamma^c{}_{ab} - \Gammaz^c{}_{ab} \right) 
 - H^c = 0.
\label{Eq:HarmConstr}
\end{equation}
Here, $H^c$ is a given vector field on $M$ and $\Gamma^c{}_{ab}$ and
$\Gammaz^c{}_{ab}$ are the Christoffel symbols corresponding to the
dynamical and background metrics, respectively. In the particular case
where $H^c = 0$ and where the background metric is the Minkowski
metric in standard Cartesian coordinates, $\Gammaz^c{}_{ab}$ vanishes,
and the condition ${\cal C}^c = 0$ reduces to the usual condition for
harmonic coordinates $\Box x^\mu = 0$ for $\mu=t,x,y,z$. However, the
advantage of the condition (\ref{Eq:HarmConstr}) is that it maintains
the covariance of the theory since ${\cal C}^c$ is the difference
between the two Christoffel symbols,
\begin{equation}
C^c{}_{ab} \equiv \Gamma^c{}_{ab} - \Gammaz^c{}_{ab}
 = \frac{1}{2} g^{cd}\left( \nablaz_{a} h_{bd} + \nablaz_{b} h_{ad} 
 - \nablaz_d h_{ab} \right),
\label{Eq:Christoffel}
\end{equation}
where $h_{ab} = g_{ab} - \gz_{ab}$ denotes the difference between the
dynamical and the background metric.

With the condition (\ref{Eq:HarmConstr}), Einstein's field equations
are equivalent to the wave system
\begin{eqnarray}
g^{cd}\nablaz_c\nablaz_d h_{ab} 
&=& 2\, g_{ef} g^{cd} C^e{}_{ac} C^f{}_{bd} 
 + 4\, C^c{}_{d(a} g_{b)e} C^e{}_{cf} g^{df}
 - 2\, g^{cd} \Rz^e{}_{cd(a} g_{b)e}
\nonumber\\
&+& 16\pi G\left( T_{ab} - \frac{1}{2} g_{ab} g^{cd} T_{cd} \right)
 +  2\,\nabla_{(a} H_{b)}\, ,
\label{Eq:EinsteinWave}
\end{eqnarray}
where $\Rz^a{}_{bcd}$ denotes the curvature tensor with respect to
$\gz_{ab}$, $T_{ab}$ the stress-energy tensor and $G$ denotes Newton's
constant. Solutions of this equation which are smooth enough imply
that the constraint variable ${\cal C}_a$ satisfies
\begin{equation}
g^{cd}\nabla_c\nabla_d {\cal C}_a = -R_a{}^b {\cal C}_b 
 - 16\pi G\nabla^b T_{ab}\; .
\label{Eq:Bianchi}
\end{equation}
Therefore, the imposition of the boundary condition $\left. {\cal C}_a
\right|_{\cal T} = 0$ implies that any smooth enough solution of
(\ref{Eq:EinsteinWave}) with initial data satisfying
\begin{displaymath}
\left. {\cal C}_a \right|_{\Sigma_0} = 0, \qquad
\left. n^a\nabla_a {\cal C}_b \right|_{\Sigma_0} = 0,
\end{displaymath}
satisfies the constraint ${\cal C}_a=0$ on $M$ provided the
stress-energy tensor is divergence free, $\nabla^b T_{ab} = 0$.

In order to formulate asymptotically nonreflecting boundary conditions
we first construct an adapted local null tetrad $\{ K^a, L^a, Q^a,
\bar{Q}^a \}$ and a radial function $r$ as in the electromagnetic
case. Notice that here these quantities are defined with respect to
the {\em dynamical} metric $g_{ab}$ and not the background metric
$\gz_{ab}$. The boundary conditions derived in section
\ref{SubSec:NRBCEinstein} are the following:
\begin{eqnarray}
\left. 
\frac{1}{r^2}\, K^a K^b K^c \nablaz_a(r^2 h_{bc}) \right|_{\cal T} 
 &=& -q_{KK}\, ,
\label{Eq:KKK}\\
\left. 
\frac{1}{r^2} K^a K^b L^c \nablaz_a(r^2 h_{bc})
 + \frac{1}{r}\gz^{bc} h_{bc} \right|_{\cal T} &=& -q_{Q\bar{Q}}\; ,
\label{Eq:KKL}\\
\left. 
\frac{1}{r^2} K^a K^b Q^c \nablaz_a(r^2 h_{bc}) \right|_{\cal T} &=& 
 -q_{KQ}\; ,
\label{Eq:KKQ}\\
\left. 
K^a Q^b Q^c \nablaz_a h_{bc} - Q^a Q^b K^c \nablaz_a h_{bc} 
\right|_{\cal T} &=& -q_{QQ}\; ,
\label{Eq:KQQ-QQK}\\
\left. \left( K^a Q^b\bar{Q}^c + L^a K^b K^c 
            - Q^a K^b\bar{Q}^c - \bar{Q}^a K^b Q^c \right)
\nablaz_a h_{bc} \right|_{\cal T} &=& \left. -2K^a H_a \right|_{\cal T}\; ,
\label{Eq:KQQbar}\\
\left. \left( K^a L^b Q^c + L^a K^b Q^c 
            - Q^a K^b L^c + \bar{Q}^a Q^b Q^c \right)
\nablaz_a h_{bc} \right|_{\cal T} &=& \left. -2 Q^a H_a \right|_{\cal T}\; ,
\label{Eq:KLQ}\\
\left. \left( K^a L^b L^c + L^a Q^b\bar{Q}^c 
            - Q^a\bar{Q}^b L^c - \bar{Q}^a Q^b L^c \right)
\nablaz_a h_{bc} \right|_{\cal T} &=& \left. -2 L^a H_a \right|_{\cal T}\; ,
\label{Eq:KLL}
\end{eqnarray}
where $q_{KK}$ and $q_{Q\bar{Q}}$ are real-valued given functions on
${\cal T}$ and $q_{KQ}$ and $q_{QQ}$ are complex-valued given
functions on ${\cal T}$. The first three equations
(\ref{Eq:KKK}),(\ref{Eq:KKL}),(\ref{Eq:KKQ}) are related to the gauge
freedom, the condition (\ref{Eq:KQQ-QQK}) controls the gravitational
radiation while the remaining conditions
(\ref{Eq:KQQbar}),(\ref{Eq:KLQ}),(\ref{Eq:KLL}) enforce the constraint
${\cal C}_a=0$ on the boundary. The evolution equation
(\ref{Eq:EinsteinWave}) has the form (\ref{Eq:WaveSystemEq}) where $E$
is the vector bundle of symmetric, covariant tensor fields on $M$ and
the boundary conditions (\ref{Eq:KKK}--\ref{Eq:KLL}) have the form
(\ref{Eq:WaveSystemBC}) where $\alpha=1$ and $c^{a\, bc}{}_{de}$ is in
upper triangular form when expressed in terms of the basis $\{ K^b
K^c,K^{(b} L^{c)},K^{(b}Q^{c)},Q^b
Q^c,Q^{(b}\bar{Q}^{c)},L^{(b}Q^{c)},L^b L^c \}$.

\section{The wave equation on a curved background}
\label{Sect:SimpleWave}

In this section we prove Theorem \ref{Thm:MainTheorem} for the case of
a single wave equation
\begin{equation}
g^{ab}\nabla_a \nabla_b\phi = S
\label{Eq:Weq}
\end{equation}
on $M = [0,T] \times \Sigma$. For simplicity, we also assume that
$g_{ab}$ and $S$ are independent of $\phi$. In this case, it is
convenient to choose $\nabla_a$ to be the Levi-Civita connection with
respect to $g_{ab}$. The IBVP consists in finding solutions of
(\ref{Eq:Weq}) subject to the initial conditions
\begin{equation}
\left. \phi \right|_{\Sigma_0} = \phi_0\; , \qquad
\left. n^b\nabla_b\phi \right|_{\Sigma_0} = \pi_0\; ,
\label{Eq:WeqInitial}
\end{equation}
where $\phi_0$ and $\pi_0$ are given functions on $\Sigma_0$, and the
boundary conditions
\begin{equation}
\left[ T^b\nabla_b\phi + \alpha N^b\nabla_b\phi \right]_{\cal T} = G,
\label{Eq:WeqBoundary}
\end{equation}
where $G$ is a given function on ${\cal T}$. Here, $n^b$ and $N^b$
denote the future-directed unit vector field to the time-slices
$\Sigma_t$ and the outward unit normal vector field to ${\cal T}$,
respectively, $T^b$ is an arbitrary future-directed time-like vector
field which is tangent to the boundary surface ${\cal T}$ and $\alpha$
is a strictly positive function on ${\cal T}$. Without loss of
generality, we assume that $T^a$ is normalized such that $g_{ab} T^a
T^b = -1$. Furthermore, by redefining $\phi$ and $S$ if necessary, we
may also assume that the boundary data $G$ vanishes identically.

In order to show well posedness for this problem, we use a geometric
reduction to a first order symmetric hyperbolic system with maximal
dissipative boundary conditions \cite{rG96,kF58,pLrP60}. First,
introducing the variables $V_a = \nabla_a\phi$, the wave equation can
be rewritten as the first order system
\begin{eqnarray}
&& \nabla_a\phi = V_a\, ,\\
&& g^{ab}\nabla_a V_b = S, \\
&& \nabla_a V_b - \nabla_b V_a = 0.
\end{eqnarray}
Next, we specify any future-directed time-like vector field $u^a$ and
contract the first and the last equation with it. This yields the
evolution system
\begin{eqnarray}
&& \pounds_u\phi = u^a V_a \equiv \Pi ,
\label{Eq:WE1}\\
&& g^{ab}\nabla_a V_b = S,
\label{Eq:WE2}\\
&& \pounds_u V_b = \nabla_b\Pi,
\label{Eq:WE3}
\end{eqnarray}
where $\pounds_u$ denotes the Lie derivative with respect to $u^a$.
This system is subject to the initial and boundary conditions
\begin{eqnarray}
&& \left. \phi \right|_{\Sigma_0} = \phi_0\; , \qquad
   \left. n^b V_b \right|_{\Sigma_0} = \pi_0\; , \qquad 
   \iota_0^* V_b = \iota_0^*\nabla_b\phi_0\; ,
\label{Eq:WEInitial}\\
&& \left[ T^b V_b + \alpha N^b V_b \right]_{\cal T} = 0 ,
\label{Eq:WEBoundary}
\end{eqnarray}
where $\iota_0: \Sigma_0 \to M$ is the inclusion map, and subject to the
constraint $C_a = 0$, where the constraint variable $C_a$ is defined
as $C_a = V_a - \nabla_a\phi$.  The evolution equations (\ref{Eq:WE1})
and (\ref{Eq:WE3}) imply that $C_a$ is Lie-dragged by the time
evolution vector field $u^a$,
\begin{displaymath}
\pounds_u C_a = 0.
\end{displaymath}
In the following, we assume that $u^a$ is pointing {\em away from the
domain} at the boundary. This implies that a solution of
(\ref{Eq:WE1},\ref{Eq:WE2},\ref{Eq:WE3}) with constraint-satisfying
initial data automatically satisfies the constraints everywhere on
$M$, and no extra boundary conditions are needed in order to ensure
that the constraint $C_a = 0$ propagates.

Still, there is a huge freedom in choosing the evolution vector field
$u^a$; different choices lead to first order evolution systems
(\ref{Eq:WE1},\ref{Eq:WE2},\ref{Eq:WE3}) which are inequivalent to
each other if the solution is off the constraint surface $C_a = 0$. In
this work we exploit this freedom in order to obtain energy estimates
which allow for an appropriate control of the fields not only in the
bulk but also on the boundary of the domain (see the estimate (\ref
{Eq:SWP}) below). In order to analyze this, following \cite{rG96} we
rewrite the evolution system (\ref{Eq:WE2},\ref{Eq:WE3}) in the form
\begin{displaymath}
{\cal A}^a{}_{bc} \nabla_a V^c 
 \equiv -u^a(\nabla_a V_b - \nabla_b V_a) + u_b \nabla_a V^a = u_b S,
\end{displaymath}
where the symbol is given by ${\cal A}^a{}_{bc} = -u^a g_{bc} +
2\delta^a{}_{(b} u_{c)}$. Since ${\cal A}^a{}_{bc}$ is symmetric in
$bc$ and since $u_a {\cal A}^a{}_{bc} = -u_a u^a g_{bc} + 2 u_b u_c$
is positive definite, the evolution system is symmetric hyperbolic.
In particular, the evolution equations imply that
\begin{displaymath}
\nabla_a ({\cal A}^a{}_{bc} V^b V^c) 
 = (\nabla_a {\cal A}^a{}_{bc}) V^b V^c + 2(u_b V^b) S.
\end{displaymath}
Integrating both sides of this equation over the manifold $M = [0,T]
\times \Sigma$ and using Gauss' theorem, one obtains\footnote{Notice
that since $n^a$ is future directed, its flow increases $t$; hence in
coordinates $(t,x^i)$ where $t$ parametrizes $[0,T]$ and $x^i$ are
local coordinates on $\Sigma$, we have $n^t > 0$ and $n_t < 0$.}
\begin{equation}
\int\limits_{\Sigma_T} n_a {\cal A}^a{}_{bc} V^b V^c  
 = \int\limits_{\Sigma_0} n_a {\cal A}^a{}_{bc} V^b V^c 
 + \int\limits_{{\cal T}} N_a {\cal A}^a{}_{bc} V^b V^c
 - \int\limits_M 
   \left[ (\nabla_a {\cal A}^a{}_{bc}) V^b V^c + 2(u_b V^b) S \right].
\label{Eq:EnergyEstimate}
\end{equation}
The following two conditions (see \cite{pLrP60}) guarantee that the
IBVP
(\ref{Eq:WE1},\ref{Eq:WE2},\ref{Eq:WE3},\ref{Eq:WEInitial},\ref{Eq:WEBoundary})
is well posed:
\begin{enumerate}
\item[(i)] $n_a {\cal A}^a{}_{bc}$ is positive definite.
\item[(ii)] For each $p\in {\cal T}$, the subspace ${\cal N}_-(p)
\subset T_p M$ consisting of the vectors $V^b(p)$ satisfying the
boundary condition (\ref{Eq:WEBoundary}) at $p$ is maximal
non-positive. This means that $N_a {\cal A}^a{}_{bc}(p) V^b(p) V^c(p)
\leq 0$ for all $V^b(p)\in {\cal N}_-(p)$ and that ${\cal N}_-(p)$
does not posses a proper extension with this property.
\end{enumerate}

For the following, we choose the time evolution vector field $u^a$
such that $u^a$ is everywhere future-directed and time-like on $M$ and
such that $u^a$ lies in the plane spanned by $T^a$ and $N^a$ at each
point of the boundary, more specifically,
\begin{displaymath}
\left. u^a \right|_{\cal T} = T^a + \delta N^a,
\end{displaymath}
with $0 < \delta < 1$ a function on ${\cal T}$. The following two
lemmas imply the satisfaction of the conditions (i) and (ii) for an
appropriate choice of $\delta$.

\begin{lemma}
\label{Lem:Positivity1}
$n_a {\cal A}^a{}_{bc}(p)$ is positive definite for all $p\in M$.
\end{lemma}

\proof Let $h_{ab} = g_{ab} + n_a n_b$ be the induced metric on
$\Sigma_t$ and expand $u_a = \mu(n_a + \bar{u}_a)$, where $\mu = -n^a
u_a$. Since $u^a$ is future-directed and time-like, $\mu > 0$ and
$\bar{u}^a\bar{u}_a < 1$. Therefore,
\begin{displaymath}
n_a {\cal A}^a{}_{bc} 
 = \mu\left( h_{bc} + n_b n_c + 2 n_{(b}\bar{u}_{c)} \right)
\end{displaymath}
is positive definite.
\qed

\begin{lemma}
\label{Lem:MD1}
Let $0 < \delta \leq \alpha(1 + \alpha^2)^{-1}$. Then, the boundary
spaces ${\cal N}_-(p)$ are maximal non-positive for all $p\in {\cal
T}$.
\end{lemma}

\proof (cf. appendix B in Ref. \cite{hKoRoSjW07}) 
Fix a point $p\in {\cal T}$, and let $V^b\in T_p M$. We have
\begin{eqnarray}
N_a {\cal A}^a{}_{bc} V^b V^c &=& 
 \left[ \delta \, T_b T_c + \delta\, N_b N_c + 2 T_{(b} N_{c)}
  - \delta\, H_{bc} \right] V^b V^c
\nonumber\\
 &=& -\delta\left[ (T^b V_b)^2 + (N^b V_b)^2 + H_{bc} V^b V^c \right]
     + 2\left[ \delta (T^b V_b)^2 + \delta (N^b V_b)^2 + (T^b V_b)(N^c V_c)
        \right],
\nonumber
\end{eqnarray}
where $H_{bc} = g_{bc} + T_b T_c - N_b N_c$ is the induced metric on
the orthogonal complement of the plane spanned by $T^b$ and
$N^b$. Eliminating the terms $(T^b V_b)$ in the second square bracket
on the right-hand side using the boundary condition
(\ref{Eq:WEBoundary}) we obtain
\begin{equation}
N_a {\cal A}^a{}_{bc} V^b V^c 
 = -\delta\left[ (T^b V_b)^2 + (N^b V_b)^2 + H_{bc} V^b V^c \right]
 + 2\left[ \delta(\alpha^2 + 1) - \alpha \right] (N^b V_b)^2.
\label{Eq:Boundary1}
\end{equation}
The last term on the right-hand side is non-positive by the assumption
of the lemma. Therefore, $N_a {\cal A}^a{}_{bc}$ is negative-definite
on the subspace of vectors $V^a$ satisfying the boundary condition.
Finally, we observe that ${\cal N}_-(p)$ is maximal since its
dimension is $d = \dim T_p M - 1$ while the symmetric bilinear form
$N_a {\cal A}^a{}_{bc}$ has signature $(1,d)$.
\qed

If we relax the assumption of homogeneous boundary data and replace
the condition (\ref{Eq:WEBoundary}) by the condition
\begin{equation}
\left[ T^b V_b + \alpha N^b V_b \right]_{\cal T} = G,
\label{Eq:WEBoundaryIH}
\end{equation}
we obtain, instead of (\ref{Eq:Boundary1}),
\begin{displaymath}
N_a{\cal A}^a{}_{bc} V^b V^c = 
 -\delta\left[ (T^b V_b)^2 + (N^b V_b)^2 + H_{bc} V^b V^c \right]
 + 2\left[ \delta(\alpha^2 + 1) - \alpha \right] (N^b V_b)^2
 + 2(1 - 2\delta\alpha)(N^b V_b) G + 2\delta\, G^2.
\end{displaymath}
Let $0 < \rho < 1$ and set $\delta = (1-\rho)\alpha(1 +
\alpha^2)^{-1}$. Then, we have (cf. appendix B in
Ref. \cite{hKoRoSjW07})
\begin{equation}
N_a{\cal A}^a{}_{bc} V^b V^c \leq
 -\delta\left[ (T^b V_b)^2 + (N^b V_b)^2 + H_{bc} V^b V^c \right]
 + \left[ 2\delta + \frac{(1-2\delta\alpha)^2}{2\alpha\rho} \right] G^2.
\label{Eq:BoundaryEstimate}
\end{equation}
Using this and the positivity of $n_a {\cal A}^a{}_{bc}$ in the
identity (\ref{Eq:EnergyEstimate}) we obtain the estimate
\begin{displaymath}
\int\limits_{\Sigma_t} n_a{\cal A}^a{}_{bc} V^b V^c
 \leq \int\limits_{\Sigma_0} n_a{\cal A}^a{}_{bc} V^b V^c
 - C_1\int\limits_{{\cal T}_t} n_a{\cal A}^a{}_{bc} V^b V^c
 + C_2\int\limits_{{\cal T}_t} G^2
 + C_3\int\limits_0^t \left[ 
   \int\limits_{\Sigma_s} n_a{\cal A}^a{}_{bc} V^b V^c
 + \int\limits_{\Sigma_s} S^2 \right] ds
\end{displaymath}
for all $0 \leq t \leq T$, where $C_1$, $C_2$ and $C_3$ are strictly
positive constants which are independent of $V^b$, and ${\cal T}_t :=
[0,t] \times \partial\Sigma$. Applying Gronwall's lemma\footnote{See,
for instance, Lemma 3.1.1 in Ref. \cite{KL-Book}} to the function
$y(t) := \int\limits_0^t \int\limits_{\Sigma_s} n_a{\cal A}^a{}_{bc}
V^b V^c ds$ we obtain from this

\begin{lemma}
Let $T > 0$. There is a constant $C = C(T) \geq 1$ such that all
smooth enough solutions to the IBVP
(\ref{Eq:WE2},\ref{Eq:WE3},\ref{Eq:WEInitial},\ref{Eq:WEBoundaryIH})
satisfy the inequality
\begin{equation}
\int\limits_{\Sigma_t} n_a{\cal A}^a{}_{bc} V^b V^c 
 + \int\limits_{{\cal T}_t} n_a{\cal A}^a{}_{bc} V^b V^c
 \leq C\left[ \int\limits_{\Sigma_0} n_a{\cal A}^a{}_{bc} V^b V^c 
 + \int\limits_{{\cal T}_t} G^2 
 + \int\limits_0^t \left(\int\limits_{\Sigma_s} S^2 \right) ds \right],
\label{Eq:SWP}
\end{equation}
for all $0 \leq t \leq T$, where ${\cal T}_t := [0,t] \times
\partial\Sigma$.
\end{lemma}

Since any solution of this problem also satisfies $u^a C_a = u^a V_a -
\pounds_u\phi = 0$, $\pounds_u C_a = 0$ and $\iota_0^* C_a = \iota_0^*
(V_a - \nabla_a\phi) = 0$, and since $u^a$ points outward from the
domain at ${\cal T}$, the constraint $C_a = 0$ is satisfied everywhere
on $M$. From this and the previous lemma, we have established: 

\begin{theorem}
The second order problem
(\ref{Eq:Weq},\ref{Eq:WeqInitial},\ref{Eq:WeqBoundary}) is strongly
well posed: given smooth initial and boundary data $\phi_0$, $\pi_0$
and $G$ satisfying the usual compatibility conditions at
$\partial\Sigma_0$, there exists a unique smooth solution satisfying
the estimate (\ref{Eq:SWP}) with $V^a$ replaced by $\nabla^a\phi$.
\end{theorem}

\begin{remark}
The important feature of the estimate (\ref{Eq:SWP}) is the second
term on the left-hand side which yields a $L^2$ boundary estimate for
the gradient of $\phi$. This estimate is obtained by choosing the time
evolution vector field $u^a$ in such a way that the boundary matrix
$N_a {\cal A}^a{}_{bc}$ is negative definite on the subspace of
vectors satisfying the boundary conditions. As we will see (Lemma
\ref{Lem:MD2} in the next section), this property is important for
systems of wave equations since it allows the coupling of the boundary
conditions through small enough terms involving first derivatives of
the fields. If, on the other hand, $u^a$ is chosen to be tangent to
the boundary, the boundary matrix has a nontrivial kernel and one does
not obtain an estimate for the gradient of $\phi$ on the boundary from
the first order system. However, this does not affect the strong
well posedness of the second order system which is independent of
$u^a$.
\end{remark}

As an example, consider the wave equation on the half-plane $\Sigma =
\Real_+ \times \Real^2$ with the flat metric $g = -dt^2 + dx^2 + dy^2
+ dz^2$. In this case, we have
\begin{displaymath}
n^a\partial_a = \partial_t\; ,\qquad
N^a\partial_a = -\partial_x\; \qquad
T^a\partial_a = \frac{1}{p}
\left( \partial_t - \beta^y\partial_y - \beta^z\partial_z \right),
\end{displaymath}
with $(\beta^y)^2 + (\beta^z)^2 < 1$ and $p := \sqrt{1 - (\beta^y)^2 -
(\beta^z)^2}$, and the boundary condition (\ref{Eq:WeqBoundary})
reduces to
\begin{equation}
\left[ \phi_t + p\alpha \phi_x - \beta^y\phi_y - \beta^z\phi_z \right]_{x=0} 
 = p G,
\label{Eq:SingleWaveBC}
\end{equation}
where $\phi_t := \partial_t\phi$ etc. Choosing $u^a = p(T^a + \delta
N^a)$ with $0 < \delta < 1$ the energy norm for this problem reads
\begin{displaymath}
\int\limits_{\Sigma_t} n_a{\cal A}^a{}_{bc} V^b V^c
 = \int\limits_0^\infty\int\limits_{-\infty}^\infty\int\limits_{-\infty}^\infty
 \left[ \phi_t^2 + \phi_x^2 + \phi_y^2 + \phi_z^2 
 + 2\phi_t\left( \delta p\phi_x + \beta^y\phi_y + \beta^z\phi_z \right)
 \right] dy\, dz\, dx.
\end{displaymath}
This is similar to the norm we used in Ref. \cite{hKoRoSjW07} for
obtaining an a priori energy estimates for the second order wave
equation with boundary condition (\ref{Eq:SingleWaveBC}).

\section{Systems of wave equations and proof of main theorem}
\label{Sect:WaveSystems}

In order to show that the system
(\ref{Eq:WaveSystemEq},\ref{Eq:WaveSystemID},\ref{Eq:WaveSystemBC})
yields a well posed IBVP, we follow the arguments given in
Sect.~\ref{Sect:SimpleWave} and reduce it to a first order symmetric
hyperbolic system with maximal dissipative boundary conditions. Let
$V_a{}^A := \nabla_a\Phi^A$, and let $u^a(p,\Phi)$ denote a
future-directed time-like vector field on $M$ such that
\begin{displaymath}
\left. u^a \right|_{\cal T} = T^a + \delta N^a,
\end{displaymath}
with $0 < \delta < 1$ a function on ${\cal T}$ to be determined. Then
(\ref{Eq:WaveSystemEq}) can be rewritten as the first order
evolution system
\begin{eqnarray}
u^a\nabla_a\Phi^A &=& u^a V_a{}^A, 
\label{Eq:WaveSystemFO1}\\
g^{ab}(\Phi)\nabla_a V_b{}^A &=& S^A(\Phi,V),
\label{Eq:WaveSystemFO2}\\
u^a\left( \nabla_a V_b{}^A - \nabla_b V_a{}^A \right)
 &=& u^a R^A{}_{Bab}\Phi^B,
\label{Eq:WaveSystemFO3}
\end{eqnarray}
where $R^A{}_{Bab}$ denotes the curvature belonging to the connection
$\nabla_a$. At this point, we stress that the connection $\nabla_a$ is
a fixed background connection on the vector bundle $E$, and not the
Levi-Civita connection belonging to the metric $g_{ab}(\Phi)$, so that
$R^A{}_{Bab}$ does not depend on $\Phi$ nor its derivatives. The
system
(\ref{Eq:WaveSystemFO1},\ref{Eq:WaveSystemFO2},\ref{Eq:WaveSystemFO3})
is subject to the constraint $C_b{}^A = 0$, where $C_b{}^A :=
\nabla_b\Phi^A - V_b{}^A$.
Eqs. (\ref{Eq:WaveSystemFO1},\ref{Eq:WaveSystemFO3}) imply that the
constraint variable $C_b{}^A$ is Lie-dragged by $u^a$:
\begin{displaymath}
\pounds_u C_b{}^A \equiv u^a\nabla_a C_b{}^A + (\nabla_b u^a) C_a{}^A = 0.
\end{displaymath}
Therefore, any smooth enough solution of the first order problem
(\ref{Eq:WaveSystemFO1},\ref{Eq:WaveSystemFO2},\ref{Eq:WaveSystemFO3})
belonging to initial data with $C_b{}^A = 0$ satisfies the constraint
$C_b{}^A = 0$ everywhere it is defined. The initial condition is
\begin{equation}
\left. \Phi^A \right|_{\Sigma_0} = \Phi^A_0\; , \qquad
\left. n^b V^A_b \right|_{\Sigma_0} = \Pi^A_0\; , \qquad 
\iota_0^* V^A_b = \iota_0^*\nabla_b\Phi^A_0\; ,
\label{Eq:WaveSystemFOID}
\end{equation}
and the boundary condition (\ref{Eq:WaveSystemBC}) reads
\begin{equation}
\left[ T^b V_b + \alpha N^b V_b\right]_{\cal T}
 = c^{a\, A}{}_B\left. V_a{}^B \right|_{\cal T} 
 + d^A{}_B\left. \Phi^B \right|_{\cal T} + G^A.
\label{Eq:WaveSystemFOBC}
\end{equation}

In order to analyze the well posedness of the first order IBVP
(\ref{Eq:WaveSystemFO1},\ref{Eq:WaveSystemFO2},
\ref{Eq:WaveSystemFO3},\ref{Eq:WaveSystemFOID},\ref{Eq:WaveSystemFOBC})
we first linearize the system by replacing the coefficients
$g_{ab}(\Phi)$, $S^A(\Phi,\nabla\Phi)$, $T^b(\Phi)$, $N^b(\Phi)$,
$\alpha(\Phi)$, $c^{a\, A}{}_B(\Phi)$, $d^A{}_B(\Phi)$ by smooth
functions $g_{ab}$, $S^A$, $T^b$, $N^b$, $\alpha$, $c^{a\, A}{}_B$,
$d^A{}_B$, respectively. Local in time well posedness for the original
quasilinear system follows by iteration from the well posedness result
for the linear system with enough differentiability\footnote{See, for
instance, \cite{jRfM874,KL-Book}.}. Next, we use a partition of unity
in order to localize the problem. With this, it is sufficient to
consider a local trivialization $\varphi: U \times \Real^N \mapsto
\pi^{-1}(U)$ of $E$ such that $\bar{U}\subset M$ is compact and
contains a portion ${\cal U}$ of the boundary ${\cal T}$. Let
$\varepsilon > 0$. According to the assumption there exists a smooth
map $J_\varepsilon: U \to GL(N,\Real), p \mapsto (J_\varepsilon(p))$
such that the transformed matrix coefficients $\tilde{c}^a :=
J_\varepsilon c^a J_\varepsilon^{-1}$ satisfy the condition
(\ref{Eq:hABCondition}) for all vector-valued one-forms $V_a$ on
${\cal U}$. Setting $h_{AB}(\varepsilon) := (J_\varepsilon^T h
J_\varepsilon)_{AB} = h_{CD} (J_\varepsilon)^C{}_A
(J_\varepsilon)^D{}_B$, we can reformulate this condition by stating
that
\begin{equation}
  h_{AB}(\varepsilon) c^{a\, A}{}_C(\Phi) c^{b\, B}{}_D(\Phi) V_a{}^C V_b{}^D 
   \leq \varepsilon h_{AB}(\varepsilon) e^{ab}(\Phi) V_a{}^A V_b{}^B\; ,
\label{Eq:hABConditionBis}
\end{equation}
for all vector-valued one-forms $V_a^A$ on ${\cal U}$. The system
(\ref{Eq:WaveSystemFO1},\ref{Eq:WaveSystemFO2},\ref{Eq:WaveSystemFO3})
can be written in the form
\begin{equation}
\left( \begin{array}{cc} -\Lambda h_{AB}(\varepsilon) u^a\nabla_a & 0 \\
 0 & h_{AB}(\varepsilon) {\cal A}^a{}_{bc}\nabla_a \end{array} \right)
\left( \begin{array}{c} \Phi^B \\ V^{c\, B} \end{array} \right)
 = {\cal S}(\Phi,V),
\label{Eq:WaveSystemFOSH}
\end{equation}
where $\Lambda > 0$ is to be determined, ${\cal A}^a{}_{bc} = -u^a
g_{bc} + 2\delta^a{}_{(b} u_{c)}$ and
\begin{displaymath}
{\cal S}(\Phi,V) = \left( \begin{array}{l}
 -\Lambda h_{AB}(\varepsilon) u^a V_a{}^B \\
  -h_{AB}(\varepsilon) R^B{}_{Cab}\Phi^C u^a 
 + h_{AB}(\varepsilon) u_b S^B(\Phi,V)
 \end{array} \right).
\end{displaymath}
Let $B(n_a;(\Psi,W),(\Phi,V))$ denote the bilinear form belonging to
the principal symbol of (\ref{Eq:WaveSystemFOSH}), that is, for an
arbitrary one-form $w_a$ on $M$ define
\begin{displaymath}
B(w_a;(\Psi,W),(\Phi,V)) := -\Lambda u^a w_a h_{AB}(\varepsilon)\Psi^A\Phi^B
 + h_{AB}(\varepsilon) w_a {\cal A}^a{}_{bc} W^{b\, A} V^{c\, B}.
\end{displaymath}
We have

\begin{lemma}
Let $\Lambda > 0$. Then, $B(n_a;(\Psi,W),(\Phi,V))$ is symmetric in
$(\Psi,W),(\Phi,V)$ and positive definite for $w_a = u_a$ and $w_a =
n_a$. Therefore, the system (\ref{Eq:WaveSystemFOSH}) is symmetric
hyperbolic.
\end{lemma}

\proof The symmetry property follows immediately from the symmetry of
$h_{AB}(\varepsilon)$ and the symmetry of ${\cal A}^a{}_{bc}$ in
$bc$. In order to check the positivity statements, let $w_a = u_a$,
$\gamma := \sqrt{-u^a u_a}$ and $\hat{u}_a := \gamma^{-1} u_a$. Since
${\cal A}^a{}_{bc} u_a = \gamma^2\left[ g_{bc} + 2\hat{u}_b\hat{u}_c
\right]$, we find
\begin{displaymath}
B(u_a;(\Phi,V),(\Phi,V)) = \gamma^2\left[
 \Lambda h_{AB}(\varepsilon) \Phi^A\Phi^B 
 + ( g_{ab} + 2\hat{u}_a\hat{u}_b ) h_{AB}(\varepsilon) V_{a\, A} V^{b\, B}
\right]
\end{displaymath}
which is manifestly positive definite. The proof that
$B(n_a;(\Phi,V),(\Phi,V))$ is positive definite is similar to the
proof of Lemma \ref{Lem:Positivity1}.
\qed

As in the previous section we obtain well posedness of the linearized
system provided we can show that each boundary space
\begin{displaymath}
{\cal N}_-(p) := \{ (\Phi,V)\in \Real^N \times \Real^{(d+1)N} : 
\left[ T^b(p) + \alpha(p) N^b(p) \right] V_b{}^A 
 = c^{a\, A}{}_B(p) V_a{}^B + d^A{}_B(p)\Phi^B \},
\qquad p\in {\cal U},
\end{displaymath}
is maximal non-positive with respect to $B(N_a;(\Phi,V),(\Phi,V))$.
This is the statement of the next lemma.

\begin{lemma}
\label{Lem:MD2}
Set $\delta := \alpha(1 + \alpha^2)^{-1}/2$ and $\kappa := 2[ 2\delta
+ (1-2\delta\alpha)^2/\alpha]^2$. Choose $\varepsilon > 0$ small
enough such that $\kappa\varepsilon < \delta$ and $\Lambda > 0$ large
enough such that $2\kappa h_{AB}(\varepsilon) d^A{}_C
d^B{}_D\Phi^C\Phi^D \leq \delta\Lambda
h_{AB}(\varepsilon)\Phi^A\Phi^B$ for all $\Phi\in\Real^N$. Then, the
boundary space ${\cal N}_-(p)$ is maximal non-positive for all $p\in
{\cal U}$.
\end{lemma}

\proof Let $p\in {\cal U}$. We have, as in the proof of Lemma
\ref{Lem:MD1},
\begin{eqnarray}
B(N_a;(\Phi,V),(\Phi,V)) 
 &=& -\Lambda u^a N_a h_{AB}(\varepsilon) \Phi^A\Phi^B
  + h_{AB}(\varepsilon) N_a {\cal A}^a{}_{bc} V^{b\, A} V^{c\, B}
\nonumber\\
 &=& -\delta h_{AB}(\varepsilon)
      \left[ (T^a T^b + N^a N^b + H^{ab}) V_a{}^A V_b{}^B
 + \Lambda \Phi^A\Phi^B \right]
\nonumber\\
 &+& 2\left[ \delta\, T^a T^b + \delta\, N^a N^b + T^a N^b \right] 
 h_{AB}(\varepsilon) V_a{}^A V_b{}^B.
\label{Eq:Bilinear1}
\end{eqnarray}
Let $(\Phi^A,V_a{}^A)\in {\cal N}(p)$. Then, $T^a V_a{}^A = -\alpha
N^a V_a{}^A + \tilde{G}^A$ with $\tilde{G}^A := c^{a\, A}{}_B V_a{}^B
+ d^A{}_B\Phi^B$, and we may use this equation in order to eliminate
the terms $(T^a V_a{}^A)$ in the second bracket on the right-hand side
of (\ref{Eq:Bilinear1}). This yields
\begin{displaymath}
B(N_a;(\Phi,V),(\Phi,V)) \leq 
 -\delta h_{AB}(\varepsilon)\left[ (T^a T^b + N^a N^b + H^{ab}) V_a{}^A V_b{}^B
 + \Lambda \Phi^A\Phi^B \right]
 + \left[ 2\delta + \frac{(1-2\delta\alpha)^2}{\alpha} \right] 
 h_{AB}(\varepsilon)\tilde{G}^A\tilde{G}^B,
\end{displaymath}
where we have set $\delta := \alpha(1 + \alpha^2)^{-1}/2$ and used the
boundary estimate (\ref{Eq:BoundaryEstimate}) with $\rho = 1/2$. Now,
\begin{eqnarray}
h_{AB}(\varepsilon)\tilde{G}^A\tilde{G}^B 
 &\leq& 2h_{AB}(\varepsilon) c^{a\, A}{}_C V_a{}^C c^{b\, B}{}_D V_b{}^D
 + 2h_{AB}(\varepsilon) d^A{}_C\Phi^C d^B{}_D\Phi^D
\nonumber\\
 &\leq& 2\varepsilon h_{AB}(\varepsilon) e^{ab} V_a{}^A V_b{}^B
  + 2 h_{AB}(\varepsilon) d^A{}_C d^B{}_D\Phi^C\Phi^D,
\end{eqnarray}
where we have used the estimate (\ref{Eq:hABConditionBis}) in the last
step. Recalling that $e^{ab} = g^{ab} + 2T^a T^b = T^a T^b + N^a N^b +
H^{ab}$ and the definition of $\kappa$ in the assumption of the lemma
we find
\begin{displaymath}
B(N_a;(\Phi,V),(\Phi,V)) \leq 
 -\delta h_{AB}(\varepsilon)\left[ e^{ab} V_a{}^A V_b{}^B
 + \Lambda \Phi^A\Phi^B \right]
 + \kappa\left[ \varepsilon h_{AB}(\varepsilon) e^{ab} V_a{}^A V_b{}^B 
   + h_{AB}(\varepsilon) d^A{}_C d^B{}_D\Phi^C\Phi^D \right].
\end{displaymath}
The non-positivity of ${\cal N}_-(p)$ now follows from the assumptions
on $\varepsilon$ and $\Lambda$. Finally, we observe that an element in
${\cal N}_-(p)$ is characterized by $N$ conditions in a
$(d+2)N$-dimensional space which implies that $\dim{\cal N}_-(p) \geq
(d+1)N$. On the other hand, from Eq. (\ref{Eq:Bilinear1}) we see that
the signature of $B(N_a; . , . )$ is given by $(N,(d+1)N)$. Therefore,
$\dim {\cal N}_-(p) = (d+1)N$ and the maximality of ${\cal N}_-(p)$
follows. \qed

\section{Boundary conditions for isolated systems}
\label{Sect:Applications}

We consider here boundary conditions for an isolated system emitting
radiation. If, for computational purposes, the evolution domain of
such a system has a finite (artificial) boundary, some artificial
boundary condition must be imposed. If one knew the correct boundary
data for the analytic problem, then in principle one could use any
boundary condition corresponding to a well posed IBVP. However, the
determination of the correct boundary data is in general a global
problem, in which the boundary data must be determined by extending
the solution to infinity either by matching to an exterior (linearized
or nonlinear) solution obtained by some other means. The matching
approach has been reviewed elsewhere~\cite{livrevccm}. Here we
consider an alternative approach in which homogeneous boundary data
can be assigned in such a way that the accuracy of the boundary
condition becomes exact in the limit that the boundary is extended to
infinity. (Such boundary conditions would also be beneficial to the
matching approach because the corresponding boundary data would be
small so that numerical or other error would also have a small
effect.) Artificial boundary conditions for an isolated radiating
system for which homogeneous data is approximately valid are commonly
called absorbing boundary conditions (see
e.g.~\cite{EnquistMajda-77,Hig86,Tre86,Bla88,Jia90,Ren96}), or
nonreflecting boundary conditions (see e.g. \cite{Hed79,giv91,kell98})
or radiation boundary conditions (see e.g. \cite{Bay80}). Such
boundary conditions are advantageous for computational use. However,
local artificial boundary condition are not perfectly nonreflecting in
general. Here, to be more precise, we consider nonreflecting boundary
conditions in the sense of boundary conditions for a well posed
problem for which homogeneous data produces no spurious reflection in
the limit that the boundary approaches an infinite sphere. The
extensive literature on improved versions of nonreflecting boundary
conditions involves higher order and nonlocal methods. Our interest
here is to investigate the optimal choice of local first order
homogeneous boundary conditions on a spherical boundary for the
constrained Maxwell and linearized Einstein problems expressed in
terms of the gauge dependent variables $A^\mu$ and $\gamma^{\mu\nu}$.
See \cite{sL05,lBoS06,lBoS07} for the construction of higher-order and
higher-accurate boundary conditions for Einstein's equations.

We base our discussion on waves from an isolated system satisfying a
system of flat space wave equations. We use Greek indices to denote
standard inertial coordinates $x^\mu =(t,x,y,z)$ in which the
components of the Minkowski metric $\eta^{\mu\nu}$ are
$\diag(-1,1,1,1)$. In the case of a scalar field $\Phi$, we thus
consider the wave equation
\begin{displaymath}
    \eta^{\alpha\beta}\partial_\alpha \partial_\beta \Phi
        = \left(-\partial_t^2+\partial_x^2+\partial_y^2+\partial_z^2 \right) 
          \Phi=S,
\end{displaymath}
where the source $S$ has compact support. Outside the source, we
assume that the solution has the form
\begin{equation}
    \Phi=\frac{f(t-r,\theta,\phi)}{r} +\frac{g(t-r,\theta,\phi)}{r^2}
    +\frac{h(t,r,\theta,\phi)}{r^3}\; ,
    \label{eq:outgoing}
\end{equation}
where $(r,\theta,\phi)$ are standard spherical coordinates and $f$,
$g$ and $h$ and their derivatives are smooth bounded functions. These
assumptions determine the exterior retarded field of a system emitting
outgoing radiation. The simplest case is the monopole radiation
\begin{displaymath}
    \Phi=\frac{f(t-r)}{r}
\end{displaymath}
which satisfies $(\partial_t+\partial_r) (r\Phi)=0$. This motivates
the use of a Sommerfeld condition
\begin{displaymath}
  \frac{1}{r}(\partial_t+\partial_r)(r\Phi)|_R= q(t,R,\theta,\phi)
\end{displaymath}
on a finite boundary $r=R$. 

The resulting Sommerfeld boundary data $q$ in the general case
(\ref{eq:outgoing}) falls off as $1/R^3$, so that a homogeneous
Sommerfeld condition introduces an error which is vanishingly small
for increasing $R$. As an example, for the dipole solution
\begin{displaymath}
    \Phi_{Dipole}=\partial_z{\frac{f(t-r)}{r}}
                 =-\left( \frac{f'(t-r)}{r} + \frac{f(t-r)}{r^2} \right)
                  \cos\theta
\end{displaymath}
we have
\begin{displaymath}
    q= \frac{f(t-r)\cos\theta}{R^3}\; .
\end{displaymath}
A homogeneous Sommerfeld condition at $r=R$ would lead to a solution
$\tilde \Phi_{Dipole}$ containing a reflected ingoing wave. For large
$R$,
\begin{displaymath}
  \tilde \Phi_{Dipole} \sim \Phi_{Dipole}
  + \kappa \frac{F(t+r-2R)\cos\theta}{r}\; ,
\end{displaymath}
where $\partial_t f(t)=F(t)$ and the reflection coefficient has
asymptotic behavior $\kappa=O(1/R^2)$. More precisely, the Fourier
mode
\begin{displaymath}
  \tilde \Phi_{Dipole}(\omega) = \partial_z\bigg (\frac{e^{i\omega (t-r)}}{r}
         + \kappa_\omega \frac{e^{i\omega (t+r-2R)}}{r} \bigg ),
\end{displaymath}
satisfies the homogeneous boundary condition
$(\partial_t+\partial_r)(r\tilde\Phi_{Dipole}(\omega)|_R=0$ with
reflection coefficient
\begin{equation}
          \kappa_\omega =\frac{1} {2\omega^2 R^2 +2i\omega R-1}
                 \sim \frac{1} {2\omega^2 R^2} .
\label{eq:skappa}
\end{equation}

Note that (59) and (62) satisfy
\begin{equation}
          \kappa \sim q R.
\label{eq:qkappa}
\end{equation}
In the case of a system of equations $\kappa$ will have $N$ components
corresponding to the number of modes generated in the reflected wave.
The boundary conditions lead to a system of simultaneous equations
relating $\kappa$ to the components of the Sommerfeld data $q$. If these
equations are nondegenerate then (\ref{eq:qkappa}) continues to hold.
However, degeneracies could conceivably lead to weaker asymptotic
falloff of $\kappa$. (It would be interesting to determine whether such
cases exist.) In any case, (\ref{eq:qkappa}) gives the optimum allowable
behavior of the reflection coefficients so that the asymptotic behavior
of the Sommerfeld data $q$ is a good indicator of the quality of the
boundary condition. This forms the basis of our investigation of the
Maxwell and linearized Einstein equations with a spherical boundary in
sections~\ref{SubSec:NRBCMaxwell} and \ref{SubSec:NRBCEinstein}.

\subsection{A plane boundary}
\label{sec:plane}

The key ideas in the above example are that (i) the Sommerfeld
condition is only satisfied exactly by waves traveling in the radial
direction and (ii) in the asymptotic limit $r \rightarrow\infty$ all
waves from an isolated system propagate in the radial direction. This
allows us to reformulate our discussion of the Sommerfeld condition by
considering a wave $\Phi$ propagating in the domain $x<0$, which is
incident on a plane boundary at $x=0$ with the boundary
condition
\begin{displaymath}
    K^\alpha\partial_\alpha \Phi|_{x=0} =0,
\end{displaymath} 
where $K^\alpha \partial_\alpha =\partial_t+\partial_x$ is the
characteristic direction determined by the outward normal to the
boundary $\partial_x$ and the time direction $\partial_t$. This
homogeneous condition is satisfied for plane waves $\Phi=G(t+k_x x+k_y
y+k_z z)$ incident on the boundary only for the single case
$(k_x,k_y,k_z)=(1,0,0)$, i.e. a plane wave propagating in the outgoing
normal direction. Plane waves in the normal direction pass through the
boundary whereas plane waves incident in other directions on the
boundary give rise to a reflected wave. We will take advantage of this
simplification of the plane wave case in discussing boundary
conditions for electromagnetic and gravitational waves. The results
then suggest how to formulate boundary conditions for an isolated
electromagnetic or gravitational system with a spherical boundary of
radius $R$, where in the limit $R\rightarrow \infty$ all radiation is
incident normally.

For the electromagnetic case, we describe the field by means of a
vector potential $A^\mu$ satisfying the Lorentz gauge
condition. Maxwell's equations in a flat spacetime with Minkowski
metric $\eta^{\mu\nu}$ then reduce to the wave equations
$$
     \eta^{\alpha\beta} \partial_\alpha \partial_\beta A^\mu =0 
$$
subject to the constraint
$$
   C:= \partial_\mu A^\mu =0
$$ introduced by the Lorentz gauge condition. This constraint keeps us
from requiring that each component of $A^\mu$ satisfy a homogeneous
Sommerfeld condition, in contrast to the scalar example. The
electromagnetic case also differs from the scalar case because of the
remaining gauge freedom allowed by the Lorentz condition.

An electromagnetic plane wave incident in the outgoing normal
direction can be described by the real part of the vector potential
\begin{displaymath}
    A_\mu=F(t-x)Q_\mu + G(t-x) K_\mu\; ,
\end{displaymath}
where $F(t-x)$ is complex, $Q^\mu=Y^\mu+iZ^\mu$ is a complex null
polarization vector, $G(t-x)$ represents gauge freedom and
$K^\mu=T^\mu +X^\mu$, in terms of the orthonormal tetrad
$(T^\mu,X^\mu,Y^\mu,Z^\mu)$ aligned with the coordinate axes
satisfying
\begin{displaymath}
     \eta_{\mu\nu}=-T_\mu T_\nu+X_\mu X_\nu+Y_\mu Y_\nu+Z_\mu Z_\nu\; .
\end{displaymath}

In order to formulate a gauge invariant boundary condition we consider
the corresponding electromagnetic field tensor
\begin{displaymath}
    F_{\mu\nu}=\partial_\mu A_\nu -\partial_\nu A_\mu
     =-F^\prime(t-x)(K_\mu Q_\nu -Q_\mu K_\nu).
\end{displaymath}
Here we adopt the notation $\partial_u F(u) =F'(u)$. For this plane
wave, all components of $F_{\mu\nu}$ satisfy
$$
        K^\mu F_{\mu\nu}=0.
$$
However, this condition rules out the possibility of a static
electric field oriented normal to the boundary. For the purpose of
formulating a boundary condition which only restricts propagating
waves it suffices to consider the weaker condition
\begin{equation}
        K^\mu Q^\nu F_{\mu\nu}=0.
\label{eq:fbc}
\end{equation}
In terms of the electric and magnetic field components tangential to
the boundary, (\ref{eq:fbc}) corresponds to the plane wave relations
${\bf E}_{tan}\cdot{\bf B}_{tan}=0$ and $|{\bf E}_{tan}|=|{\bf
B}_{tan}|$, with the corresponding Poynting vector in the outward
normal direction.

We can incorporate (\ref{eq:fbc}) into the following homogeneous
Sommerfeld boundary conditions for the vector potential:
\begin{eqnarray}
        &K^\nu K^\mu\partial_\mu  A_\nu = 0 ,
      \label{eq:Kbc}\\
        &Q^\nu K^\mu \partial_\mu A_\nu 
            =  K^\nu Q^\mu \partial_\mu  A_\nu.
\label{eq:Qbc}
\end{eqnarray}
The remaining boundary condition can be expressed in Sommerfeld form
by rewriting the constraint as
\begin{equation}
    C=\frac{1}{2}\left( -L^\nu  K^\mu -K^\nu L^\mu
     + Q^\nu \bar Q^\mu  +\bar Q^\nu  Q^\mu \right) \partial_\mu  A_\nu =0,
\label{eq:cbc}
\end{equation}
where $L^\mu=T^\mu-X^\mu$. Here $(K^\mu, L^\mu, Q^\mu)$
form a null tetrad according to the conventions
\begin{equation}
         \eta_{\mu\nu}= -K_{(\mu}L_{\nu)}+ Q_{(\mu}\bar Q_{\nu)} .
\label{eq:ntetrad}
\end{equation}
We assume throughout the following that the spin transformation
freedom $Q^\mu \rightarrow e^{i\alpha}Q^\mu$ has been restricted
according to $K^\mu\partial_\mu \alpha =0$. The Sommerfeld boundary
conditions (\ref{eq:Kbc}), (\ref{eq:Qbc}) and (\ref{eq:cbc}) have the
required hierarchical, upper triangular form for a well posed IBVP,
see Lemma \ref{Lem:LyapunovTrick}.

For the purpose of extending this approach to the gravitational case,
we write the linearized Einstein vacuum equations in the form
\begin{equation}
   \eta^{\alpha\beta}\partial_\alpha \partial_\beta \gamma^{\mu\nu}=0
\label{eq:lingw}
\end{equation}
subject to the harmonic constraints
\begin{equation} 
   C^\nu:= -\partial_\mu \gamma^{\mu\nu} =0.
\label{eq:lingc}
\end{equation}
Here, to linearized accuracy, we set
$\sqrt{-g}g^{\mu\nu}=\eta^{\mu\nu}+ \gamma^{\mu\nu}$ so that
$\gamma_{\mu\nu}$ represents the perturbation of the densitized
metric. (Indices of linearized objects are raised and lowered with the
Minkowski metric.)

A plane gravitational wave incident on the boundary in the outgoing
normal direction is given by
\begin{displaymath}
      \gamma^{\mu\nu}=F(t-x)Q^\mu Q^\nu +K^{(\mu}\xi^{\nu)}(t-x)
      -\frac{1}{2}\eta^{\mu\nu} K^\alpha\xi_\alpha(t-x),
\end{displaymath}
where $\xi^\nu(t-x)$ describes the gauge freedom.

In order to formulate a boundary condition with gauge invariant
meaning analogous to (\ref{eq:fbc}) in the Maxwell case, we consider
the linearized curvature tensor
\begin{eqnarray}
    -2 R_{\mu\nu\rho\sigma} &=& \partial_\rho \partial_\nu \gamma_{\mu\sigma}
       -\partial_\sigma \partial_\nu \gamma_{\mu\rho}
      -\partial_\rho \partial_\mu \gamma_{\nu\sigma}
        +\partial_\sigma \partial_\mu \gamma_{\nu\rho}
      -\frac{1}{2}(\eta_{\mu\sigma}\partial_\rho \partial_\nu \gamma
                   -\eta_{\mu\rho}\partial_\sigma \partial_\nu \gamma
                   -\eta_{\nu\sigma}\partial_\rho \partial_\mu \gamma
                   +\eta_{\nu\rho}\partial_\sigma \partial_\mu \gamma)
\nonumber\\
  &=& 4F''(t-x) K_{[\mu} Q_{\nu]} Q_{[\rho} K_{\sigma]}\; .
\end{eqnarray}
Plane wave boundary conditions on the curvature tensor could be
imposed by requiring that the Newman-Penrose component $\Psi_0=K^\mu
Q^\nu Q^\rho K^\sigma R_{\mu\nu\rho\sigma}$ vanish on the
boundary. (See~\cite{hFgN99} for a discussion of the appropriateness
of this boundary condition.) However, this requirement involves second
derivatives in the normal direction when expressed in terms of
$\gamma_{\mu\nu}$. Instead, we require $\Psi:=K^\mu Q^\nu Q^\rho
T^\sigma R_{\mu\nu\rho\sigma}=0$ on the boundary.  The condition
$\Psi=0$ is equivalent to $\Psi_0=0$ if the Ricci component
$R_{\mu\nu}Q^\mu Q^\nu=0$, e.g.  if the vacuum Einstein equations are
satisfied.

A straightforward calculation leads to
\begin{eqnarray}
 -2\Psi&=&  K^\mu Q^\nu Q^\rho T^\sigma (
     \partial_\rho \partial_\nu \gamma_{\mu\sigma}
       -\partial_\sigma \partial_\nu \gamma_{\mu\rho}
      -\partial_\rho \partial_\mu \gamma_{\nu\sigma}
        +\partial_\sigma \partial_\mu \gamma_{\nu\rho} ) 
       +\frac{1}{2}Q^\nu Q^\rho\partial_\nu \partial_\rho \gamma
     \nonumber \\
   &=& K^\mu Q^\nu Q^\rho T^\sigma ( 
            -\partial_\sigma \partial_\nu \gamma_{\mu\rho}
      -\partial_\rho \partial_\mu \gamma_{\nu\sigma}
        +\partial_\sigma \partial_\mu \gamma_{\nu\rho} )
        +\frac{1}{2}(K^\mu K^\sigma +Q^\mu \bar Q^\sigma)
         Q^\nu Q^\rho \partial_\nu \partial_\rho \gamma_{\mu\sigma} 
     \nonumber \\
  &=& Q^\nu \partial_\nu \bigg ( 
     \frac{1}{2}(K^\mu K^\sigma +Q^\mu \bar Q^\sigma )
            Q^\rho \partial_\rho \gamma_{\mu\sigma}
     -K^\mu  Q^\rho T^\sigma  \partial_\mu \gamma_{\sigma\rho} \bigg ) 
         \nonumber \\
    &+&T^\sigma\partial_\sigma \bigg ( 
        - K^\mu Q^\nu Q^\rho \partial_\nu \gamma_{\mu\rho} 
        + K^\mu Q^\nu Q^\rho \partial_\mu \gamma_{\nu\rho}  \bigg ).
\label{eq:psi1}	
\end{eqnarray}

Thus, besides containing no second derivatives normal to the boundary,
the condition $\Psi=0$ can be reduced to two first order conditions by
factoring out the $Q^\nu\partial_\nu$ and $T^\sigma\partial_\sigma$
derivatives in (\ref{eq:psi1}) which are tangential to the
boundary. There are many ways this can be done. In order to obtain
first order conditions which fit into a hierarchy of Sommerfeld
conditions, we modify (\ref{eq:psi1}) according to the steps
\begin{eqnarray}
   -2\Psi&=&Q^\nu \partial_\nu \bigg ( 
     \frac{1}{2}(K^\mu K^\sigma  +Q^\mu \bar Q^\sigma)
         Q^\rho \partial_\rho \gamma_{\mu\sigma}
        -\frac{1}{2}K^\rho  Q^\mu L^\sigma \partial_\rho \gamma_{\mu\sigma}
     -\frac{1}{2}K^\mu  Q^\rho K^\sigma  \partial_\mu \gamma_{\sigma\rho} 
                \bigg ) 
         \nonumber \\
    &+&T^\sigma\partial_\sigma \bigg ( 
        - K^\mu Q^\nu Q^\rho \partial_\nu \gamma_{\mu\rho} 
        + K^\mu Q^\nu Q^\rho \partial_\mu \gamma_{\nu\rho}  \bigg )
         \\
   &=&\frac{1}{2}Q^\nu \partial_\nu \bigg ( 
  (K^\mu K^\sigma  Q^\rho + Q^\mu K^\sigma  L^\rho -Q^\mu Q^\sigma \bar Q^\rho)
         \partial_\rho \gamma_{\mu\sigma} -2Q^\mu C_\mu
     -K^\mu  Q^\rho K^\sigma  \partial_\mu \gamma_{\sigma\rho} 
                \bigg ) 
         \nonumber \\
    &+&T^\sigma\partial_\sigma \bigg ( 
        - K^\mu Q^\nu Q^\rho \partial_\nu \gamma_{\mu\rho}
       + K^\mu Q^\nu Q^\rho \partial_\mu \gamma_{\nu\rho}  \bigg ) 
       \\
     &=&\frac{1}{2}Q^\nu \partial_\nu \bigg ( 
  (K^\mu K^\sigma  Q^\rho -Q^\mu Q^\sigma \bar Q^\rho)
         \partial_\rho \gamma_{\mu\sigma} 
     -2K^\mu  Q^\rho K^\sigma  \partial_\mu \gamma_{\sigma\rho} 
            -2Q^\mu C_\mu    \bigg ) 
         \nonumber \\
    &+&T^\sigma\partial_\sigma \bigg ( 
       K^\mu Q^\nu Q^\rho \partial_\mu \gamma_{\nu\rho}  \bigg ) .
       \label{eq:psi}
\end{eqnarray}
Thus since the derivatives $Q^\nu \partial_\nu$ and $T^\nu
\partial_\nu$ are tangential to the boundary, we can enforce $\Psi=0$
on the boundary through the first order boundary conditions
\begin{eqnarray}
     & Q^\alpha Q^\beta K^\mu\partial_\mu\gamma_{\alpha\beta} = 0 ,    \\
    &K^\alpha Q^\beta K^\mu\partial_\mu \gamma_{\alpha\beta} - 
        \frac{1}{2}K^\alpha K^\beta  Q^\mu\partial_\mu \gamma_{\alpha\beta}    
     +\frac{1}{2}Q^\alpha Q^\beta \bar Q^\mu \partial_\mu \gamma_{\alpha\beta} 
     =0. 
 \end{eqnarray}
These two boundary conditions can then be included in a hierarchical
set of Sommerfeld boundary conditions, according to the example
\begin{eqnarray}
     &  K^\alpha K^\beta K^\mu\partial_\mu \gamma_{\alpha\beta} = 0,
      \label{eq:KKbc}\\
     & Q^\alpha Q^\beta K^\mu\partial_\mu \gamma_{\alpha\beta} = 0,        
      \label{eq:QQbc}\\
  &Q^\alpha \bar Q^\beta K^\mu\partial_\mu \gamma_{\alpha\beta} = 0,        
      \label{eq:QbQbc}\\
  &  K^\alpha Q^\beta K^\mu\partial_\mu \gamma_{\alpha\beta} - 
        \frac{1}{2} K^\alpha K^\beta Q^\mu\partial_\mu\gamma_{\alpha\beta}
   +\frac{1}{2}Q^\alpha Q^\beta \bar Q^\mu \partial_\mu \gamma_{\alpha\beta}  
     =0   \label{eq:KQbc}.
\end{eqnarray}
The constraints $C_\rho=0$, which determine the remaining boundary
conditions, can be cast in the Sommerfeld form
\begin{displaymath}
 C_\rho=\frac{1}{2}\bigg  ( L^\nu  K^\mu +K^\nu L^\mu -\bar Q^\nu Q^\mu 
       - Q^\nu \bar Q^\mu\bigg ) \partial_\mu \gamma_{\nu\rho} =0,
\end{displaymath}
which can also be incorporated into the hierarchy. 

However, there are many alternative possibilities to (\ref{eq:KKbc}) -
(\ref{eq:KQbc}) which preserve the hierarchical Sommerfeld structure
and lead to a well posed IBVP.  In the absence of a clear geometric
approach, we next examine the boundary conditions appropriate to an
isolated system by considering the resulting reflection off a
spherical boundary.

\subsection{Application to Maxwell fields with a spherical boundary}
\label{SubSec:NRBCMaxwell}

In the case of a general retarded solution for a massless scalar wave
equation, we found that a Sommerfeld boundary condition on a spherical
boundary of radius $R$ required data $q=O(1/R^3)$. Homogeneous
Sommerfeld data gave rise to an ingoing wave with reflection
coefficient $\kappa=O(1/R^2)$, as in (\ref{eq:skappa}). This is the
best that can be achieved with a local first order homogeneous
boundary condition on a spherical boundary. We now investigate the
corresponding result for the constrained Maxwell equations expressed
in terms of a vector potential $A^\mu$.

In doing so, we associate spherical coordinates $(r,x^A)$,
$x^A=(\theta,\phi)$, in a standard way with the Cartesian coordinates
$x^i=(x,y,z)$, e.g.  $z=r\cos\theta$. As in (\ref{eq:ntetrad}) we
introduce a null tetrad $(K^\mu, L^\mu, Q^\mu)$ adapted to the
boundary, where now $K^\mu\partial_\mu=\partial_t+\partial_r$,
$L^\mu\partial_\mu=\partial_t-\partial_r$ and we fix the spin-rotation
freedom in the complex null vector vector $Q^\mu=(0,Q^i)$ by setting
\begin{equation}
   Q^i = \frac {\partial x^i}{\partial x^A} Q^A,
   \label{eq:Qi}
\end{equation}
where 
\begin{displaymath}
    Q^A= \left( Q^{\theta},Q^\phi \right)
       = \frac{1}{r}\left( 1,\frac{i}{\sin\theta} \right).
\end{displaymath}
We describe outgoing waves in terms of the retarded time $u=t-r$.

In order to investigate the vector potential describing the exterior
radiation field emitted by an isolated system we introduce a Hertz
potential with the symmetry
\begin{displaymath}
        H^{\mu\nu}=H^{[\mu\nu]}+\frac{1}{4}\eta^{\mu\nu}H.
\end{displaymath}
Then the vector potential
\begin{displaymath}
     A^\mu=\partial_\nu H^{\mu\nu}
\end{displaymath}
satisfies the Lorentz gauge condition and generates a solution of
Maxwell's equations provided the Hertz potential satisfies the wave
equation. The trace $H$ represents pure gauge freedom.

We consider outgoing dipole waves oriented with the $z$-axis. Other
dipole waves can be generated by a rotation. Higher multipole waves
can be generated by taking spatial derivatives.

The choice $H=Z^\alpha\partial_\alpha \frac{F(u)}{r}$,
$H^{[\mu\nu]}=0$ gives rise to the dipole gauge wave
\begin{displaymath}
 A_\mu=\bigg(\frac{F''(u)}{r}+\frac{F'(u)}{r^2}\bigg)\cos\theta K_\mu 
   +\bigg(\frac{2F'(u)}{r^2}+\frac{3F(u)}{r^3}\bigg)\cos\theta\partial_\mu r
         -\bigg(\frac{F'(u)}{r^2}+\frac{F(u)}{r^3}\bigg)Z_\mu
\end{displaymath} 
with components 
\begin{eqnarray}
       K^\mu A_\mu=\bigg(\frac{F'(u)}{r^2}+\frac{2F(u)}{r^3}\bigg)cos\theta, 
          \nonumber \\
       Q^\mu A_\mu=\bigg(\frac{F'(u)}{r^2}+\frac{F(u)}{r^3}\bigg)\sin\theta .
       \label{eq:gdip}
\end{eqnarray}
In appendix \ref{sec:app} we give some useful formulae underlying the
calculation leading to (\ref{eq:gdip}) and the following results.

The choice $ H^{\mu\nu}=(T^\mu Z^\nu-Z^\mu T^\nu)\frac{f(u)}{r}$ gives
rise to a dipole electromagnetic wave
\begin{displaymath}
    A_\mu=-\bigg(\frac{f'(u)}{r}+\frac{f(u)}{r^2}\bigg) T_\mu \cos\theta
           -\frac{f'(u)}{r}Z_\mu
\end{displaymath}
with components 
\begin{eqnarray}
       A^\mu K_\mu=\frac{f(u)}{r^2}cos\theta, \nonumber \\
       A^\mu Q_\mu=\frac{f'(u)}{r}\sin\theta .
       \label{eq:dip}
\end{eqnarray}

The choice $ H^{\mu\nu}=(X^\mu Y^\nu-Y^\mu X^\nu)\frac{f(u)}{r}$ gives
rise to a dipole electromagnetic wave with the dual polarization
\begin{displaymath}
    A_\mu=-\bigg(\frac{f'(u)}{r}+\frac{f(u)}{r^2}\bigg) 
    \bigg(\frac{yX_\mu}{r}-\frac{xY_\mu}{r}\bigg)
\end{displaymath}
with components 
\begin{eqnarray}
       A^\mu K_\mu=0, \nonumber \\
       A^\mu Q_\mu=i\bigg(\frac{f'(u)}{r}+\frac{f(u)}{r^2}\bigg)\sin\theta.
       \label{eq:ddip}
\end{eqnarray}

We wish to formulate boundary conditions which generalize the
Sommerfeld hierarchy (\ref{eq:Kbc}) and (\ref{eq:Qbc}) to a spherical
boundary of radius $R$ in a way which minimizes reflection. By
inspection of (\ref{eq:gdip}), (\ref{eq:dip}) and (\ref{eq:ddip}), we
consider the choice
\begin{eqnarray}
     \frac{1}{r^2}K^\mu\partial_\mu(r^2 K^\nu A_\nu)=q_K \; ,
    \label{eq:maxbc1} \\
    \frac{1}{r}K^\mu\partial_\mu(r Q^\nu A_\nu)
       - Q^\mu\partial_\mu( K^\nu A_\nu) =q_Q,
       \label{eq:maxbc2}
\end{eqnarray}
chosen to minimize the asymptotic behavior of the Sommerfeld data. As
before, the constraint determines the remaining boundary condition as
part of the Sommerfeld hierarchy. 

For the dipole gauge wave (\ref{eq:gdip}), 
\begin{displaymath}
   q_K=-\frac{2F(u)\cos\theta}{R^4}\; , \qquad q_Q = 0;
\end{displaymath}
for the dipole electromagnetic wave (\ref{eq:dip}), 
\begin{displaymath}
   q_K=0, \qquad q_Q = \frac{f(u)}{R^3} \sin\theta;
\end{displaymath}
and for the dual dipole electromagnetic wave  (\ref{eq:ddip})
\begin{displaymath}
   q_K=0, \qquad q_Q = \frac{-if(u)}{R^3} \sin\theta .
\end{displaymath}
Overall this implies $q_K=O(1/R^4)$ and $q_Q=O(1/R^3)$. We have checked
that homogeneous Sommerfeld data leads to reflection coefficients with
overall behavior $\kappa=O(1/R^2)$ in accordance with (\ref{eq:qkappa}).

Note that the relations (\ref{eq:KK}) and (\ref{eq:rQK}) allow us to
express (\ref{eq:maxbc1}) and (\ref{eq:maxbc2}) in the form
\begin{eqnarray}
     \frac{1}{r^2} K^\nu K^\mu\partial_\mu(r^2 A_\nu)=q_K\; , 
  \label{eq:mxbc1} \\
   Q^\nu K^\mu\partial_\mu A_\nu
       -K^\nu Q^\mu\partial_\mu  A_\nu =q_Q,
       \label{eq:mxbc2}
\end{eqnarray}
which correspond to (\ref{Eq:MaxwellBC1}) and (\ref{Eq:MaxwellBC2}) when
$\partial_\mu$ is generalized to the connection $\nabla_a$ in a curved
space background. Here (\ref{eq:mxbc2}) is equivalent to the gauge
invariant condition
\begin{equation}
      Q^\nu  K^\mu F_{\mu\nu} =q_Q.
\label{eq:Phi}	 
\end{equation}

\subsection{Application to linearized gravitational fields with a spherical boundary}
\label{SubSec:NRBCEinstein}

The gravitational case is more complicated than the electromagnetic
case because the geometry of the boundary is coupled with the boundary
condition.  Additionally, there are no gauge invariant quantities,
analogous to (\ref{eq:Phi}) in the electromagnetic case, on which to
base first order boundary conditions. We begin with a discussion of
how to adapt to a curved boundary the first order version of the
$\Psi$ boundary condition given in Sect.~\ref{sec:plane} for a plane
boundary.

In the nonlinear treatment of a curved boundary with unit outer normal
$N^a$ we can decompose the metric according to
\begin{displaymath}
       g_{ab}=\tau_{ab}+N_a N_b\; ,
\end{displaymath}
where $\tau_{ab}$ is the metric intrinsic to the time-like
boundary. Let $D_a$ denote the covariant derivative associated with
$\tau_{ab}$. The extrinsic curvature of the boundary is
\begin{displaymath}
    N_{ab}=\tau_a{}^c \nabla_c N_b\; .
\end{displaymath}
We complete an orthonormal basis by setting
\begin{displaymath}
          \tau_{ab}= -T_a T_b +Q_{(a} \bar Q_{b)}
\end{displaymath}
in terms of a time-like vector $T^a$ and complex null vector
$Q^a$ tangent to the boundary.

We decompose $\Psi:=K^a Q^b Q^c T^d R_{abcd}=\Psi_T +\Psi_N$ and the
Weyl component $\Psi_0 = K^a Q^b Q^c K^d R_{abcd}=\Psi_T +\Psi_N +
2\Psi_{TN}$, where $K^a =T^a +N^a$ and
\begin{eqnarray}
  \Psi_T    &=& T^a Q^b Q^c T^d R_{abcd}\; ,\\
  \Psi_N    &=& N^a Q^b Q^c T^d R_{abcd}\; ,\\
  \Psi_{TN} &=& T^a Q^b Q^c N^d R_{abcd}\; .
\end{eqnarray}
When the vacuum Einstein equations are satisfied the Riemann curvature
tensor may be replaced by the Weyl tensor whose symmetry implies
$\Psi_{TN} = 0$. Therefore, in this case, $\Psi=0$ implies the
vanishing of the Newman-Penrose Weyl component $\Psi_0=0$.

A short calculation gives the embedding formulae
\begin{displaymath}
           \Psi_N = Q^b Q^c T^d 
                (D_d N_{bc}-D_b N_{cd} )         
\end{displaymath}
and 
\begin{displaymath}
           \Psi_T =T^a Q^b Q^c T^d \left( {}^{(3)} R_{abcd}
                  -N_{ac}N_{bd}+N_{bc}N_{ad} \right),
\end{displaymath}
where ${}^{(3)} R_{abcd}$ is the intrinsic curvature to
the boundary, i.e.
\begin{displaymath}
   T^a Q^b Q^c T^d {}^{(3)} R_{abcd}
            =Q^b Q^c T^d (D_d D_c -D_c D_d) T_b\; .
\end{displaymath}
(These are the embedding equations for the Cauchy problem corrected
for the space-like character of the normal to the boundary.)

We now apply these results to a spherical boundary $r=R$ in linearized
theory off a Minkowski background, i.e
$g_{\mu\nu}=\eta_{\mu\nu}+\epsilon h_{\mu\nu}$ in standard inertial
coordinates $x^\mu$, where $\epsilon$ is the linearization parameter.
We choose $T_\mu =\partial_\mu t +O(\epsilon)$ and $N_\mu
=\partial_\mu r + O(\epsilon)$. Then $D_\mu T_\nu =O(\epsilon)$ and
$N_{\mu\nu}=R^{-1} Q_{\mu\nu} + O(\epsilon)$, where $Q_{\mu\nu
}=Q_{(\mu} \bar Q_{\nu)}$ is the metric of a 2-sphere of radius $R$.
We choose the basis to satisfy $T^\mu D_\mu T_\nu =0$ and $T^\mu D_\mu
Q_\nu =0$, so that
\begin{displaymath}
  \Psi_T = T^\mu Q^\nu Q^\rho T^\sigma {}^{(3)} R_{\mu\nu\rho\sigma}
         +O(\epsilon^2)
	 = T^\sigma D_\sigma (Q^\nu Q^\rho D_\rho T_\nu)+O(\epsilon^2)
\end{displaymath}
and
\begin{displaymath}
    \Psi_N = T^\sigma D_\sigma (Q^\nu Q^\rho N_{\rho\nu})
          -Q^\rho D_\rho (Q^\nu T^\sigma N_{\sigma\nu})
     +\frac{1}{2} Q^\rho (D_\rho Q_\mu)\bar Q^\mu Q^\nu T^\sigma N_{\sigma\nu}
	  +\frac{1}{R} Q^\nu Q^\rho D_\rho T_\nu
         +O(\epsilon^2).
\end{displaymath}

Thus the boundary conditions
\begin{eqnarray}
         Q^\nu Q^\rho (N_{\rho\nu} +D_\rho T_\nu) &=& 0, \nonumber \\
	 Q^\nu T^\rho N_{\rho\nu} &=&0,
	 \label{eq:gbc} 
\end{eqnarray}
imply to linearized accuracy that
\begin{equation}
    \Psi= \frac {1}{R} Q^\nu Q^\rho D_\rho T_\nu\; .
    \label{eq:asympsi}
\end{equation}
This gives a geometric formulation of the first differential order
version of the requirement that $\Psi \rightarrow 0$ in the asymptotic
limit $R \rightarrow \infty$. However, $\Psi_0=O(1/R^5)$ in an
asymptotically flat space-time, whereas (\ref{eq:asympsi}) leads to
$\Psi=O(1/R^2)$. This is an indication that the boundary conditions
(\ref{eq:gbc}) might lead to more reflection than desirable. Can this
be remedied by the introduction of, say, lower order terms in the
boundary conditions?  We investigate this question in the context of a
well posed IBVP based upon the harmonic version of the linearized
Einstein equations (\ref{eq:lingw}) and (\ref{eq:lingc}), where
$\gamma^{\mu\nu}=-h^{\mu\nu}+\frac{1}{2}\eta^{\mu\nu}h$.

For this purpose, we now consider linearized outgoing waves in the
harmonic gauge which are incident on a spherical boundary. We model
our discussion on the Maxwell case by using the gravitational analogue
of a Hertz potential $H^{\mu\alpha\nu\beta}$
\cite{bergsachs,bergboard}, which has the symmetries
\begin{displaymath}
  H^{\mu\alpha\nu\beta}=  H^{[\mu\alpha]\nu\beta}=H^{\mu\alpha[\nu\beta]}
      =H^{\nu\beta\mu\alpha}
\end{displaymath}
and satisfies the flat space wave equation
\begin{displaymath}
  \partial^\sigma \partial_\sigma H^{\mu\alpha\nu\beta}= 0.
\end{displaymath}
Then the densitized metric perturbation
\begin{displaymath}
        \gamma^{\mu\nu} =\partial_\alpha \partial_\beta H^{\mu\alpha\nu\beta}. 
\end{displaymath}
satisfies the linearized Einstein equations in the harmonic gauge.
Outgoing waves can be generated from the potential
\begin{displaymath}
    H^{\mu\alpha\nu\beta} =\frac{f^{\mu\alpha\nu\beta}(u)}{r}\; ,
\end{displaymath}
and its spatial derivatives.

The incidence of such an outgoing wave on a boundary $r=R$ leads to
reflection, with the asymptotic falloff of the reflection coefficients
depending upon the choice of boundary conditions. We limit our
calculation of reflection coefficients to the case of outgoing
quadrupole waves, which can be obtained from the Hertz potential
\begin{equation}
    H^{\mu\alpha\nu\beta}= K^{\mu\alpha\nu\beta}\frac{f(u)}{r}\; ,
\label{eq:Khertz}
\end{equation}
where $ K^{\mu\alpha\nu\beta}$ is a constant tensor.  (All higher
multipoles can be constructed by taking spatial derivatives.)
$K^{\mu\alpha\nu\beta}$ has 21 independent components. However, the
choice $K^{\mu\alpha\nu\beta}=\epsilon^{\mu\alpha\nu\beta}$ leads to
$\gamma^{\mu\nu}=0$ so there are only 20 independent waves. These can
be further reduced to pure gauge waves, corresponding to the trace
terms in $K^{\mu\alpha\nu\beta}$, e.g.
$K^{\mu\alpha\nu\beta}=\eta^{\alpha\nu}\eta^{\beta\mu}
-\eta^{\mu\nu}\eta^{\alpha\beta}$ leads to a monopole gauge
wave. Linearized gravitational waves arise from the trace-free part of
$K^{\mu\alpha\nu\beta}$.  There are ten independent quadrupole
gravitational waves, corresponding to spherical harmonics with
$(\ell=2,-2\le m\le 2)$ in the two independent polarization
states. The other ten independent potentials comprise two monopole
gauge waves, three dipole gauge waves and five quadrupole gauge waves,
for which the linearized Riemann tensor vanishes. It suffices to
consider the following examples of waves with quadrupole dependence
aligned with the $z$-axis. Other quadrupole waves can be obtained by
rotation and have similar asymptotic behavior. Reflection coefficients
from the other monopole and dipole gauge waves are smaller and provide
no further useful information. The Hertz potential (\ref{eq:Khertz})
gives rise to the perturbation
\begin{displaymath}
      \gamma^{\mu\nu} =K^{\mu\alpha\nu\beta}
    \partial_\alpha \partial_\beta \frac{f(u)}{r}\; .
\end{displaymath}
Appendix \ref{sec:app} lists useful formula for the calculations
underlying the following results.

\subsubsection{Quadrupole-monopole gauge wave}

The Hertz potential
\begin{displaymath}
  H^{\mu\alpha\nu\beta}= \bigg( Z^\mu\eta^{\alpha\nu}Z^\beta
   +Z^\nu\eta^{\beta\mu}Z^\alpha -Z^\mu\eta^{\alpha\beta}Z^\nu 
   -Z^\beta\eta^{\nu\mu}Z^\alpha 
    \bigg)\frac {f(u)}{r}
\end{displaymath}
gives rise to a combination monopole-quadrupole gauge wave
with components
\begin{eqnarray}
 Q^\alpha Q^\beta \gamma_{\alpha\beta} &=&-2 \bigg(
      \frac{f'(u)}{r^2}+\frac{f(u)}{r^3}\bigg) \sin^2\theta,  \nonumber\\
 Q^\alpha \bar Q^\beta \gamma_{\alpha\beta} &=&
  -2\bigg ( \frac{f''(u)}{r}+\frac{2f'(u)}{r^2}+\frac{2f(u)}{r^3}
     \bigg)\cos^2\theta,  \nonumber\\  
  K^\alpha Q^\beta \gamma_{\alpha\beta} &=& -\frac{f(u)}{r^3}
          \sin\theta\cos\theta,       
  \label{eq:mqgauge} \\
  K^\alpha K^\beta \gamma_{\alpha\beta} &=& 2 \bigg(
       \frac{f'(u)}{r^2}+\frac{2f(u)}{r^3}\bigg) \cos^2\theta, \nonumber \\
   \gamma &=&- \frac{2f''(u)}{r} \cos^2\theta
      +2\bigg(\frac{f'(u)}{r^2}+\frac{f(u)}{r^3}
     \bigg)(1-3\cos^2\theta). \nonumber	  
\end{eqnarray}
Here the $\sin^2\theta$ dependence of the spin-weight 2 component
$Q^\alpha Q^\beta \gamma_{\alpha\beta}$ is a pure ${}_2 Y_{20}$
spin-weighted spherical harmonic; the $\sin\theta\cos\theta$
dependence of the spin-weight 1 component $K^\alpha Q^\beta
\gamma_{\alpha\beta}$ is a pure ${}_1 Y_{20}$ harmonic; and
the remaining spin-weight 0 components are mixtures of $Y_{00}$ and
$Y_{20}$.

\subsubsection{Quadrupole gravitational wave}

The trace-free Hertz potential
\begin{equation}
  H^{\mu\alpha\nu\beta}= \bigg(
      (T^\mu Z^\alpha-Z^\mu T^\alpha) (X^\nu Y^\beta-Y^\nu X^\beta)
     + (X^\mu Y^\alpha-Y^\mu X^\alpha) (T^\nu Z^\beta-Z^\nu T^\beta)
     \bigg)\frac{f(u)}{r}
     \label{eq:qgrwave}
\end{equation}
gives rise to a perturbation with $\gamma=0$ and components
\begin{eqnarray}
  Q^\alpha Q^\beta \gamma_{\alpha\beta} &=&2i\sin^2\theta
             \bigg( \frac{f''(u)}{r}+\frac{f'(u)}{r^2} \bigg ),  \nonumber \\
 Q^\alpha \bar Q^\beta \gamma_{\alpha\beta} &=&0,
 \label{eq:qgrav}  \\  
  K^\alpha Q^\beta \gamma_{\alpha\beta} &=& 
   i\cos\theta\sin\theta \bigg(\frac{2f'(u)} {r^2} 
          +\frac{3f(u)} {r^3}\bigg),  \nonumber \\
  K^\alpha K^\beta \gamma_{\alpha\beta} &=& 0,	\nonumber 
\end{eqnarray}
which have spin-weighted $\ell=2$, $m=0$ dependence.

\subsubsection{Dual quadrupole gravitational wave}

The trace-free Hertz potential 
\begin{displaymath}
  H^{\mu\alpha\nu\beta}= \bigg(
      (T^\mu Z^\alpha-Z^\mu T^\alpha) (T^\nu Z^\beta-Z^\nu T^\beta)
     -(X^\mu Y^\alpha-Y^\mu X^\alpha) (X^\nu Y^\beta-Y^\nu X^\beta)
     + \frac{1}{3}(\eta^{\mu\nu}\eta^{\alpha\beta}
       -\eta^{\mu\beta}\eta^{\nu\alpha})
     \bigg)\frac{f(u)}{r},
\end{displaymath}
obtained from the dual of (\ref{eq:qgrwave}),
gives gives rise to a perturbation with $\gamma=0$ and components
\begin{eqnarray}
  Q^\alpha Q^\beta \gamma_{\alpha\beta} &=&2\sin^2\theta
             \bigg( \frac{f''(u)}{r}+\frac{f'(u)}{r^2}+\frac{f(u)}{r^3}\bigg ),
\nonumber \\
 Q^\alpha \bar Q^\beta \gamma_{\alpha\beta} &=& 4(\cos^2\theta-\frac{1}{3})
  \bigg( \frac{f'(u)}{r^2}+\frac{f(u)}{r^3}\bigg ),
\label{eq:dqgrav} \\  
  K^\alpha Q^\beta \gamma_{\alpha\beta} &=& 
   \cos\theta\sin\theta \bigg(\frac{2f'(u)} {r^2}+\frac{f(u)} {r^3}\bigg),
 \nonumber \\
  K^\alpha K^\beta \gamma_{\alpha\beta} &=&
           2(\cos^2\theta-\frac{1}{3})\frac{f(u)} {r^3}\; ,
  \nonumber  
\end{eqnarray}
which have spin-weighted $\ell=2$, $m=0$ dependence.

\subsubsection{Sommerfeld-type boundary conditions}

Sommerfeld boundary conditions consistent with a well posed harmonic
IBVP have wide freedom regarding (i) partial derivative terms
consistent with the hierarchical upper triangular structure of the
boundary condition and (ii) lower differential order terms. Here we
consider three choices of of boundary conditions and compare their
reflection coefficients. One basic idea common to these choices has
already be used in the scalar and Maxwell cases, i.e by inspecting the
asymptotic behavior of the waves (\ref{eq:mqgauge}), (\ref{eq:qgrav})
and (\ref{eq:dqgrav}) we use the property $K^\alpha \partial_\alpha
f(u) =0$ to introduce the appropriate powers of $r$ that lead to the
smallest asymptotic behavior in the resulting Sommerfeld data.

Our first choice of boundary conditions is the mathematically simplest
choice
\begin{eqnarray}
   \frac{1}{r^2}K^\alpha K^\beta K^\mu \partial_\mu (r^2  \gamma_{\alpha\beta})
           &=& q_{KK}\; ,
 \label{eq:simpKK}\\
  \frac{1}{r} Q^\alpha Q^\beta K^\mu\partial_\mu (r  \gamma_{\alpha\beta} )
            &=& q_{QQ}\; ,
 \label{eq:simpQQ} \\
   \frac{1}{r}Q^\alpha \bar Q^\beta K^\mu\partial_\mu(r \gamma_{\alpha\beta})
          &=& q_{Q\bar Q}\; ,
 \label{eq:simpQbQ} \\
   \frac{1}{r^2}K^\alpha Q^\beta K^\mu\partial_\mu (r^2 \gamma_{\alpha\beta} ) 
	   &=& q_{KQ}\; . \label{eq:simpKQ}
\end{eqnarray}
This was the choice adopted in numerical tests verifying the stability of
the harmonic IBVP with a plane boundary \cite{mBhKjW07}. The powers of
$r$ in (\ref{eq:simpKK})-(\ref{eq:simpKQ}) are based upon the leading
asymptotic behavior of the components for the gauge wave
(\ref{eq:mqgauge}) and the gravitational waves (\ref{eq:qgrav}) and
(\ref{eq:dqgrav}). These choices lead to boundary data with the
asymptotic behavior
\begin{eqnarray}
    q_{KK} &\sim & \frac{f(u)}{R^4}\; ,  \nonumber \\
    q_{QQ}  &\sim & \frac{f'(u)}{R^3}\; ,    \nonumber\\
    q_{Q\bar Q} &\sim& \frac{f'(u)}{R^3}\; ,  \nonumber\\
    q_{KQ} &\sim & \frac{f(u)}{R^4}\; .  \nonumber
\end{eqnarray}
Thus the behavior of $q_{QQ}$ and $q_{Q\bar Q}$ imply that the
resulting reflection coefficients have overall asymptotic dependence
no weaker than $\kappa=O(1/R^2)$.

Our second choice, which is partially suggested by the electromagnetic
case (\ref{eq:maxbc2}) and leads to weaker reflection, consists of the
modifications
\begin{eqnarray}
  \frac{1}{r^2}K^\alpha K^\beta K^\mu \partial_\mu (r^2\gamma_{\alpha\beta})
           &=& q_{KK}\; ,
 \label{eq:modasymKK}\\
  \frac{1}{r^2} K^\alpha Q^\beta K^\mu\partial_\mu(r^2 \gamma_{\alpha\beta} )         &=& q_{KQ}\; ,
 \label{eq:modasymKQ}  \\
   \frac{1}{r^2}Q^\alpha\bar Q^\beta
       K^\mu\partial_\mu(r^2\gamma_{\alpha\beta})
      -\frac{\gamma}{r}   &=& q_{Q\bar Q}\; ,
 \label{eq:modasymQbQ} \\
    Q^\alpha Q^\beta K^\mu\partial_\mu  \gamma_{\alpha\beta}
     - Q^{\alpha}K^\beta Q^\mu \partial_\mu \gamma_{\alpha\beta} &=& q_{QQ}\; .
\label{eq:modasymQQ}
\end{eqnarray}
Now $q_{..}\sim f(u)/R^4$ for both gravitational quadrupole waves. For
the gauge waves, $q_{Q\bar Q} \sim f'(u)/R^3$. Using the
Regge-Wheeler-Zerilli perturbative formulation and the metric
reconstruction method described in \cite{oSmT01p} we have
independently checked that this leads to reflection coefficients
$\kappa=O(1/R^3)$ for the gravitational waves and $\kappa=O(1/R^2)$
for the gauge waves in accord with (\ref{eq:qkappa}). After replacing
$\gamma_{\mu\nu}=-h_{\mu\nu}+\frac{h}{2} \eta_{\mu\nu}$ and
identifying $\partial_\mu$ with the connection $\nablaz_{a}$ of the
background metric $\gz_{ab}$,
(\ref{eq:modasymKK})-(\ref{eq:modasymQQ}) correspond to the boundary
conditions (\ref{Eq:KKK})-(\ref{Eq:KQQ-QQK}) discussed in
Sect.~\ref{SubSec:Harmbg}.

Our third choice of boundary conditions, motivated by the first order
version of the $\Psi_0$ boundary condition (\ref{eq:KQbc}), is
\begin{eqnarray}
  K^\mu \partial_\mu (r^2 K^\alpha K^\beta \gamma_{\alpha\beta}) &=& q_{KK}\; ,
 \label{eq:rKKbcy}\\
  K^\mu\partial_\mu (r Q^\alpha Q^\beta \gamma_{\alpha\beta} ) &=& q_{QQ}\; ,
 \label{eq:rQQbcy}\\
  K^\mu\partial_\mu(r Q^\alpha \bar Q^\beta \gamma_{\alpha\beta})&=& q_{Q\bar Q}  \; ,
      \label{eq:rQbQbcy}\\
   \frac{1}{r^2}K^\mu\partial_\mu (r^2 K^\alpha Q^\beta \gamma_{\alpha\beta})
     &-&
       \frac{1}{2} Q^\mu\partial_\mu K^\alpha K^\beta \gamma_{\alpha\beta}    
     +\frac{1}{2} \bar Q^\mu  Q^\alpha Q^\beta\partial_\mu\gamma_{\alpha\beta}
	   = q_{KQ}\; .
	      \label{eq:rKQbcy}
\end{eqnarray}
However, for the gravitational quadrupole wave (\ref{eq:qgrav}), this
leads to $q_{KQ}\sim f''(u)/R^2$ and so it results in much stronger
reflection than the first two choices. Thus, as might have been
anticipated by the discussion following (\ref{eq:asympsi}), the first
order version of the $\Psi$ boundary condition is not as effective as
(\ref{eq:modasymQQ})-(\ref{eq:modasymKK}) in the case of a spherical
boundary.

\section{Conclusion}
\label{Sect:Conclusions}

We have considered the IBVP for a coupled system of quasilinear wave
equations and established (local in time) well posedness for a large
class of boundary conditions. In particular, this allows for the
formulation of a well posed IBVP for quasilinear wave systems in the
presence of constraints on finite domains with artificial,
nonreflecting boundaries. Therefore, we anticipate that our results
will have application to a wide range of problems in computational
physics. Furthermore, since our proof is based on a reduction to a
symmetric hyperbolic system with maximal dissipative boundary
conditions, it also lays the path for constructing stable finite
difference discretizations for such systems.

Our work has been motivated by the importance of the computation of
gravitational waves from the inspiral and merger of binary black
holes, which has enjoyed some recent success
\cite{fP05,mCcLpMyZ06,jBjCdCmKjvM06b,jGetal07,bSdPlRjTjW07}. At
present, however, none of the simulations of the binary black hole
problem have been based upon a well posed IBVP. The closest example is
the harmonic approach of the Caltech-Cornell group
\cite{lLmSlKrOoR06,oRlLmS07,hPdBlKlLgLmS07} which incorporates the
freezing $\Psi_0$ boundary condition in second order form and has been
shown to be well posed in the generalized sense in the high frequency
limit \cite{mRoRoS07}.

Our results have potential application to improving the binary black
hole simulations. However, many of these simulations are carried out
using the BSSN formulation \cite{mStN95,tBsS99} of Einstein's
equations, which differs appreciably from the harmonic formulation
considered here. Although our results constitute a complete analytic
treatment of the IBVP for the harmonic formulation of Einstein's
equations, the extension to the BSSN formulation is not immediately
evident. For this purpose, it would be useful to reformulate the
boundary data for the harmonic problem in terms of the intrinsic
geometry and extrinsic curvature of the boundary, as has been done for
the initial data for the Cauchy problem. Such a geometric
reformulation remains an outstanding problem.

\begin{acknowledgments}
The work of O. R. was supported in part by CONICET, SECYT-UNC and NSF
Grant INT0204937 to Louisiana State University. The work of O. S. was
supported in part by grant CIC 4.19 to Universidad Michoacana, PROMEP
UMICH-PTC-195 from SEP Mexico and CONACyT grant No. 61173. The work of
J. W. was supported by NSF grant PH-0553597 to the University of
Pittsburgh. During the course of this research we have profited from
many discussions with H. Friedrich.
\end{acknowledgments}

\appendix
\section{Some useful formulae}
\label{sec:app}
Here we give a short summary of the formulae and conventions underlying the
calculational results of Sec's.~\ref{SubSec:NRBCMaxwell}
and \ref{SubSec:NRBCEinstein}. We have
\begin{equation}
      \partial_\alpha f(u) = -f'(u)K_\alpha, \quad u=t-r, \quad 
         K^\alpha \partial_\alpha K_\beta =0 .
\label{eq:KK}
\end{equation}
so that 
\begin{equation}
      \partial_\alpha \partial_\beta \frac{f(u)}{r} = 
        \frac{f''(u)}{r}K_\alpha K_\beta +
	 \frac{f'(u)}{r^2}(K_\alpha r_\beta + r_\alpha K_\beta)
	 +\frac{2f(u)}{r^3}r_\alpha r_\beta
	 -(\frac{f'(u)}{r}+\frac{f(u)}{r^2})r_{\alpha\beta}
\end{equation}
and
\begin{equation}
    K^\mu\partial_\mu  \partial_\alpha \partial_\beta \frac{f(u)}{r} = 
      -\frac{f''(u)}{r^2}K_\alpha K_\beta 
	- \frac{2f'(u)}{r^3}(K_\alpha r_\beta + r_\alpha K_\beta)
     -\frac{6f(u)}{r^4}r_\alpha r_\beta
	 +(\frac{2f'(u)}{r^2}+\frac{3f(u)}{r^3})r_{\alpha\beta}\; ,
\end{equation}
where $r_\alpha:=\partial_\alpha r$ and
$r_{\alpha\beta}:=\partial_\alpha\partial_\beta r$.
The spatial components are
\begin{equation}
   r_i=\frac{x_i}{r}=(\sin\theta\cos\phi,\sin\theta\sin\phi,cos\theta),
       \quad r_{ij}=
            \frac{\delta_{ij}}{r}-\frac{x_i x_j}{r^3}\; .
\end{equation}

Our conventions for the polarization dyad give rise to the
Cartesian components
\begin{equation}
       (Q^x,Q^y,Q^z) =(\cos\theta\cos\phi-i\sin\phi,
             \cos\theta\sin\phi+i\cos\phi,-\sin\theta),
\end{equation}
which satisfy
\begin{equation}
        (Q^x)^2+ (Q^y)^2 = -\sin^2\theta, \quad
     Q^x \frac{y}{r}-Q^y \frac{x}{r} = -i\sin\theta, \quad
     Q^x \frac{y}{r}+Q^y \frac{x}{r} =\sin\theta \bigg(
         2\cos\theta\cos\phi\sin\phi +i(\cos^2\phi-\sin^2\phi) \bigg)
\end{equation}
and
\begin{equation}
        Q^j r_{ij} = \frac{Q_i}{r}\; , \qquad
        Q^j \partial_j Q^i = \frac{\cot\theta}{r}Q^i, \qquad
    Q^j \partial_j \bar Q^i = -\frac{\cot\theta}{r}\bar Q^i -\frac{2r^j}{r}\; .
\end{equation}
From these follow the necessary commutation relations such as
\begin{equation}
        [rQ^\mu \partial_\mu, K^\nu\partial_\nu] = 0.
\label{eq:rQK}
\end{equation}

\bibliography{refs}

\end{document}